\documentclass[12pt]{article}
\pdfoutput=1

\usepackage{slashed}
\usepackage{color, verbatim}
\usepackage{latexsym}
\usepackage{amsmath}
\usepackage{amssymb}
\usepackage{arydshln}
\usepackage[dvipsnames]{xcolor}
\usepackage{cite}
\usepackage{graphicx}
\graphicspath{{figs/}}
\usepackage{bm}
\usepackage{adjustbox}
\usepackage{booktabs}
\usepackage{multirow}
\usepackage{rotating}

\makeatletter
\@addtoreset{equation}{section}
\g@addto@macro\bfseries{\boldmath}
\newcommand\Label[1]{&\refstepcounter{equation}(\mathrm{\theequation})\ltx@label{#1}&}
\makeatother

\renewcommand{\theequation}{\thesection.\arabic{equation}}

\newcommand{\vtext}[1]{\begin{sideways}\small{#1}\end{sideways}}

\DeclareMathOperator{\tr}{tr}

\allowdisplaybreaks

\begin{document}
\thispagestyle{empty}

\vspace*{0.8cm}

\begin{center}
{\Large\sc One-loop matching in the SMEFT\\[0.4cm] 
extended with a sterile neutrino}

\vspace{0.8cm}

\textbf{Mikael Chala$^{\,a}$ and Arsenii Titov$^{\,b}$}\\
\vspace{1cm}
{\em {$^a$CAFPE and Departamento de F\'isica Te\'orica y del Cosmos,
Universidad de Granada, E--18071 Granada, Spain}}\\[0.2cm]
{\em {$^b$Dipartimento di Fisica e Astronomia ``G. Galilei'', Universit\`a degli Studi di Padova \\
and INFN, Sezione di Padova, Via Francesco Marzolo 8, I--35131 Padova, Italy}}\\[0.2cm]
\vspace{0.5cm}
\end{center}

\begin{abstract}
\noindent
We study the phenomenology of the simplest renormalisable model that, at low energy, leads to the effective field theory of the Standard Model extended with right-handed neutrinos ($\nu$SMEFT). Our aim is twofold. First, to contextualise new collider signatures in models with sterile neutrinos so far studied only using the bottom-up approach. And second and more important, to provide a thorough example of one-loop matching in the diagrammatic approach, of which other matching techniques and automatic tools can benefit for cross-checks. As byproducts of this work, we provide for the first time: \textit{(i)} a complete off-shell basis for the $\nu$SMEFT and explicit relations between operators linked by equations of motion; \textit{(ii)} a complete basis for the low-energy effective field theory ($\nu$LEFT) and the tree-level matching onto the $\nu$SMEFT; \textit{(iii)} partial one-loop anomalous dimensions in the $\nu$LEFT. This way, our work comprises a new step forward towards the systematisation of one-loop computations in effective field theories, especially if the SM neutrinos are Dirac.
\end{abstract}

\newpage

\tableofcontents

\section{Introduction}
%
Effective field theories (EFTs) are being used to describe the effects
of new heavy particles at low energy in terms of operators 
of dimension higher than four.
A well acknowledged advantage of this approach is its generality.
The only model dependence resides
in the light degrees of freedom out of which the EFT is built. 
(The symmetries of the EFT are in principle also debatable, 
but by now the group of gauge symmetries 
$SU(3)_c\times SU(2)_L\times U(1)_Y$ is well established.)
Among other aspects, this choice depends crucially on the
nature of neutrinos. If neutrinos are Majorana,
the simplest assumption is that the infrared (IR) comprises only
the Standard Model (SM) fields. The resulting EFT, known as 
SMEFT~\cite{Buchmuller:1985jz,Grzadkowski:2010es}, has been
extensively studied in the recent years; see Ref.~\cite{Brivio:2017vri} 
for a fresh review. 
If neutrinos are Dirac,
the low-energy sector has to be extended with right-handed~(RH)
singlet fermions. The corresponding EFT 
is referred to as $\nu$SMEFT~\cite{delAguila:2008ir,Liao:2016qyd}.
It has also been applied to the case in which the
new RH neutrinos are themselves Majorana, as predicted
in numerous models. 
(See Ref.~\cite{Biondini:2013xua} for an EFT 
for non-relativistic Majorana neutrinos.)
With the same spirit, other EFTs have considered also new scalars
in the IR; see \textit{e.g.} 
Refs.~\cite{Franceschini:2016gxv,Gripaios:2016xuo,Anisha:2019nzx}.

A common feature of all these EFTs is that they predict new
processes that are completely absent in the renormalisable SM. 
Many of these processes have not been studied yet experimentally. 
These include, among others, rare decays of the top quark
such as $t\to\ell^+\ell^- j$~\cite{Durieux:2014xla,Chala:2018agk}, $t\to b\overline{b} j$~\cite{Banerjee:2018fsx,Chala:2018agk} or the non-resonant $t\to b\ell^+ + E_T^\text{miss}$~\cite{Alcaide:2019pnf} as well as
rare decays of the Higgs boson including $h\to \ell^+\ell^-+4j$~\cite{Caputo:2017pit}, 
$h\to\gamma(\gamma)+E_T^\text{miss}$~\cite{Butterworth:2019iff}. 
(In Ref.~\cite{Bischer:2019ttk}  
the constraints on the $\nu$SMEFT operators 
arising from low-energy experiments have been derived.) 
However, this bottom-up approach is not without drawbacks.
Most importantly, operators other than those triggering the
signals of interest are generally also present (with correlated
coefficients) in concrete ultraviolet (UV) models; 
some of them being very constrained. 
Likewise, it is hard to prioritise one search over others.

It is therefore desirable that searches motivated 
by pure EFT inspection
are also supported by realistic UV models~%
\footnote{We are well aware that ``realistic" is an arguable concept. 
Here we adopt the notion that a ``realistic'' UV model should involve less free parameters than the EFT 
(which in turn restricts the number of new independent heavy fields),
and that there should not be large cancellations between different couplings of similar size.}.
This exercise requires matching UV models to the EFT, generally
at one loop (at which several of the most interesting and/or
dangerous operators appear often). 
In the usual diagrammatic approach,
this process consists of computing tens of
one-light-particle-irreducible off-shell amplitudes in both the
UV and the EFT. This is a very demanding task that in turn
requires knowledge of a full off-shell basis of EFT operators
(only those linked by algebraic identities and integration by
parts being removed) and their relations by equations of motion.
If low energy ($E\ll v$, with $v\sim 246$ GeV 
being the Higgs vacuum expectation value) 
observables are to be computed,
then the corresponding EFT in the electroweak (EW) symmetry 
broken phase must be also known, 
as well as its matching to the aforementioned operators.
Renormalisation group evolution (RGE) of the Wilson coefficients 
in both EFTs might be also needed.

While several of these points have been already addressed in the
SMEFT~%
\footnote{The first complete set of dimension-six operators was obtained in Ref.~\cite{Buchmuller:1985jz}. Several of them were shown to be related by equations of motion in Ref.~\cite{Grzadkowski:2010es}. The corresponding EFT below the EW symmetry breaking~(EWSB) scale, known as LEFT, was worked out in Ref.~\cite{Jenkins:2017jig}; the tree-level matching of the SMEFT onto the LEFT was also provided in the same article. This computation has been recently performed at one loop in Ref.~\cite{Dekens:2019ept}. Finally, the RGE of the SMEFT and LEFT operators was presented in Refs.~\cite{Jenkins:2013zja,Jenkins:2013wua,Alonso:2013hga} and Ref.~\cite{Jenkins:2017dyc}, respectively.}, 
very little is known about the $\nu$SMEFT beyond a full (on-shell) basis of up to dimension-seven operators~\cite{delAguila:2008ir,Aparici:2009fh,Bhattacharya:2015vja,Liao:2016qyd}. Moreover, while
new techniques~\cite{Henning:2014wua,Henning:2016lyp,Ellis:2016enq,Fuentes-Martin:2016uol,Zhang:2016pja,Ellis:2017jns} 
and tools~\cite{delAguila:2016zcb,Criado:2017khh,Bakshi:2018ics,Brivio:2019irc} 
for one-loop matching are also being developed, 
a severe obstacle for progress in this respect 
is precisely the lack of explicit one-loop matching
computations to which compare to in the literature~\cite{Brivio:2019irc}. (To the best of our knowledge, partial examples of one-loop
matching have been only provided for the SM extended with a real scalar singlet~\cite{Boggia:2016asg,Ellis:2017jns,Jiang:2018pbd}, with a charged scalar singlet~\cite{Bilenky:1993bt}, with some colourless EW multiplets for very particular parameters~\cite{Henning:2014wua} and with a vector-like quark singlet~\cite{delAguila:2016zcb}.)

In light of the discussion above, in this paper we consider 
a simple UV model whose EFT description is the $\nu$SMEFT, 
for which we provide a full off-shell basis 
and relations between different operators 
by equations of motion; see Section~\ref{sec:model}.
In Section~\ref{sec:matching} we perform 
the actual one-loop matching using the diagrammatic approach. 
We provide mathematical tools used and details of loop computations 
in Appendix~\ref{app:tools} and in Appendices~\ref{app:UV} and 
~\ref{app:dim4}, respectively.
In Section~\ref{sec:pheno}  we study the phenomenology
of the resulting EFT (with operators with correlated Wilson
coefficients, as they depend on only 
a very small number of UV couplings),
both in the Majorana and in the Dirac cases, and highlight 
the importance of performing new Higgs searches at the LHC. 
To this aim, we also rely on a full on-shell basis 
of the EFT below the EW scale and its matching onto the $\nu$SMEFT, 
as well as on partial RGE, 
all of which we provide in Appendix~\ref{app:left}.

\section{Model and effective description}
\label{sec:model}
%
We consider the SM extended with a light RH fermionic singlet $N$, as well as two heavy vector-like fermions $X_E\sim 
(\mathbf{1}, \mathbf{2})_{1/2}$, $X_N\sim (\mathbf{1}, \mathbf{1})_1$ and a heavy
singly-charged scalar $\varphi\sim (\mathbf{1},\mathbf{1})_{-1}$. 
The numbers within parentheses and the subindex indicate the 
representations of $(SU(3)_c, SU(2)_L)$ and the hypercharge $Y$, respectively.
Relatively heavy vector-like fermions and/or charged 
scalars and (one or more) sterile neutrinos 
are present in a number of models 
motivated either phenomenologically 
(\textit{e.g.} by the persistent discrepancy between the measured value 
of the muon anomalous magnetic moment 
and the corresponding SM prediction~\cite{Kannike:2011ng,Dermisek:2013gta,deJesus:2020upp}) 
or theoretically (\textit{e.g.} in models assuming 
left-right symmetry~\cite{Joshi:1991yn,Guadagnoli:2011id,Mohapatra:2014qva},  
grand unification~\cite{Fonseca:2015aoa} or 
compositeness~\cite{Chala:2012af,Cacciapaglia:2018avr}~---~in this latter 
case vector-like fermions are strictly  required by the partial compositeness 
paradigm~\cite{Kaplan:1991dc}.)

We assume CP and baryon number conservation, 
while lepton number and lepton flavour are only broken by the small 
(potentially vanishing) $N$ mass; $N$ is assumed to couple only to the electron 
(or to the muon; this choice does not alter our phenomenological results). 
Moreover, we assume that the heavy fields are odd under a $\mathbb{Z}_2$ symmetry under which all SM fields 
as well as $N$ are even.

We denote by $e, u, d$ the RH leptons and quarks; 
and by $L, Q$ the left-handed counterparts. 
We name the gluon and the EW gauge bosons by $G$ and $W, B$, 
respectively. Let us call the Higgs doublet by 
$H = [G^+, (h+iG^0)/\sqrt{2}]$ and $\tilde{H} = i\sigma_2 H^*$,  
with $\sigma_I$, $I=1,2,3$, being the Pauli matrices. 
The Lagrangian of this model reads:
\begin{equation}
\mathcal{L} = \mathcal{L}_{SM+N}+\mathcal{L}_\mathrm{heavy}\,,
\end{equation}
with
\begin{align}
 \mathcal{L}_{SM+N} &= -\frac{1}{4}G_{\mu\nu}^A G^{A \mu\nu} -\frac{1}{4} W_{\mu\nu}^I W^{I \mu\nu} -\frac{1}{4}B_{\mu\nu}B^{\mu\nu} \nonumber\\
 &\phantom{{}={}}+ \left(D_\mu H\right)^\dagger \left(D^\mu H\right) + \mu_H^2 H^\dagger H -\frac{1}{2}\lambda_H \left(H^\dagger H\right)^2\nonumber\\
 &\phantom{{}={}}+i\left(\overline{Q}\slashed{D} Q + \overline{u}\slashed{D} u + \overline{d}\slashed{D} d + \overline{L}\slashed{D} L +\overline{e}\slashed{D} e +\overline{N}\slashed{D}N\right) \nonumber\\
 &\phantom{{}={}}-\left[\frac{1}{2} m_N \overline{N^c}N + \overline{Q} Y_d H d + \overline{Q} Y_u \tilde{H}u + \overline{L}Y_e H e + \overline{L} Y_N\tilde{H} N + \mathrm{h.c.}\right],
\end{align}
and
\begin{align}
 \mathcal{L}_\mathrm{heavy} &= \overline{X_E}\left(i\slashed{D}-M_{X_E}\right)X_E + \overline{X_N}\left(i\slashed{D}-M_{X_N}\right)X_N \nonumber\\
 &\phantom{{}={}}+\left(D_\mu \varphi\right)^\ast \left(D^\mu \varphi\right)
 -M_\varphi^2\varphi^\ast\varphi 
 -\lambda_{\varphi\varphi}\left(\varphi^*\varphi\right)^2
 -\lambda_{\varphi H} \left(\varphi^\ast \varphi\right) \left(H^\dagger H\right) \nonumber\\
 &\phantom{{}={}} + \left[g_X \overline{X_E} \tilde{H} X_N + g_L \overline{X_E}\varphi^* L + 
g_N\overline{X_N}\varphi^* N + \text{h.c.}\right].
\end{align}
Our conventions for the covariant derivative of a colour singlet
field $\phi$ and for the EW field strength tensors are
\begin{equation}
 D_\mu\phi = \left(\partial_\mu - i g T^I W^I_\mu - ig' Y B_\mu\right)\phi\,, \\
 \end{equation} 
 \begin{equation}
 W_{\mu\nu}^I = \partial_\mu W_\nu^I-\partial_\nu W_\mu^I + g \varepsilon^{IJK} W_\mu^J W_\nu^K\,, \qquad
 B_{\mu\nu} = \partial_\mu B_\nu - \partial_\nu B_\mu\,,
\end{equation}
where $T_I = \sigma_I/2$ are the $SU(2)$ generators.

This model features a number of interesting properties. 
\textit{(i)}~Because of the $\mathbb{Z}_2$ symmetry, if the heavy 
particles are integrated out, 
no effective operators arise at tree level. (Note also that this symmetry 
turns the neutral component of $X_E$ into a dark matter candidate, provided 
some mechanism at a higher scale~---~which does not modify the results below~---~is invoked
to avoid direct detection constraints; we do not elaborate on this aspect of 
the phenomenology though.)
\textit{(ii)}~Because of this, it can be very easily shown that in the IR 
only tree-level amplitudes are to be computed while matching at one loop~%
\footnote{Indeed, any one-loop amplitude in the UV and in the EFT would 
read $\mathcal{M}_{UV}\sim g_{UV}/(4\pi)^2$ and $\mathcal{M}_{EFT}\sim 
\alpha_{EFT}[1 + g_{EFT}/(4\pi)^2]$, respectively. Matching $\mathcal{M}_{UV} = 
\mathcal{M}_{EFT}$ implies therefore $$\alpha_{EFT}\sim \frac{g_{UV}}{(4\pi)^2} 
\left[1-\frac{g_{EFT}}{(4\pi)^2}\right] = 
\frac{g_{UV}}{(4\pi)^2}+\mathcal{O}\left\lbrace\frac{1}{(4\pi)^4}
\right\rbrace.$$ The last term in the right-hand side of the equation is 
formally of the same order as two-loop corrections and hence negligible.}. 
(Actually, it can be shown that no loops need to be computed in the 
EFT even if tree level operators are present, but the proof is more elaborated; 
see Ref.~\cite{Manohar:2018aog}.) 
\textit{(iii)}~For the very same reason, UV corrections to light 
field propagators can be neglected~\cite{delAguila:2016zcb}. 
\textit{(iv)}~Likewise, for all practical purposes in the process of matching, any heavy renormalised mass $M$ 
(evaluated at a scale $\mu$ equal to the physical mass) 
can be identified with the physical mass itself.
Finally, for $m_N\neq 0$, this model features also the decay $N\to\nu\gamma$. Any other model fulfilling the 
aforementioned properties necessarily involves a larger number of degrees of freedom.
\begin{table}[t]
\renewcommand{\arraystretch}{1.5}
 \centering
 \adjustbox{width=\textwidth}{
 \begin{tabular}{|c|c|c|c}\hline
  $0-$Higgs & $1-$Higgs & $2-$Higgs \\
  \hline
  \textcolor{gray}{$\mathcal{O}_{DN}^1=\overline{N} \partial^2 \slashed{\partial} 
N$} & $\mathcal{O}_{NB} = \overline{L}\sigma^{\mu\nu} 
N \tilde{H} 
B_{\mu\nu}$, $\mathcal{O}_{NW}=\overline{L}\sigma^{\mu\nu} 
N\sigma_I\tilde{H} W_{\mu\nu}^I$ & $\mathcal{O}_{HN} = \overline{N}\gamma^\mu 
N (H^\dagger i D_\mu 
H)$  \\
\textcolor{gray}{$\mathcal{O}_{DN}^2= i \tilde B_{\mu\nu} 
(\overline{N}\gamma^\mu \partial^\nu N)$} & \textcolor{gray}{$\mathcal{O}_{LN}^1=\overline{L} N D^2\tilde{H}$ }, \textcolor{gray}{$\mathcal{O}_{LN}^2=\overline{L} \partial_\mu N D^\mu 
\tilde{H}$ } & \textcolor{gray}{$\mathcal{O}_{NN}^2=\overline{N} i\slashed{\partial} N(H^\dagger 
H)$} \\
\textcolor{gray}{$\mathcal{O}_{DN}^3=\partial^\nu B_{\mu\nu} (\overline{N}\gamma^\mu 
N)$} & \textcolor{gray}{$\mathcal{O}_{LN}^3=i\overline{L}\sigma^{\mu\nu} \partial_\mu N D_\nu 
\tilde{H}$ }, \textcolor{gray}{$\mathcal{O}_{LN}^4=\overline{L} (\partial^2 N)\tilde{H}$ } & $\mathcal{O}_{HNe} = \overline{N}\gamma^\mu e (\tilde{H}^\dagger iD_\mu H)$\\
  \hline
  \multicolumn{3}{|c|}{$3-$Higgs: $\mathcal{O}_{LNH} = \overline{L}\tilde{H} N (H^\dagger H)$}\\
  \hline
 \end{tabular}
 }
  \caption{\it Relevant bosonic operators. 
  The h.c.~is implied when needed. For example, $\mathcal{O}_{DN}^1 = \overline{N}\partial^2\slashed{\partial}N +\text{h.c.}$; therefore all Wilson coefficients are real.}
  \label{tab:bosonic}
\end{table}
\begin{table}[t]
\vspace{1cm}
\renewcommand{\arraystretch}{1.5}
\centering
\begin{tabular}{|c c c|}
\hline
\multirow{3}{*}{\vtext{RRRR}}&\multicolumn{2}{c|}{$\mathcal{O}_{NN}=(\overline{N}\gamma_\mu N)(\overline{N}\gamma^\mu N)$} \\
&${\cal O}_{eN}=(\overline{e}\gamma_\mu e)(\overline{N}\gamma^\mu N)$&${\cal O}_{uN}=(\overline{u}\gamma_\mu u)(\overline{N}\gamma^\mu N)$\\
&${\cal O}_{dN}=(\overline{d}\gamma_\mu d)(\overline{N}\gamma^\mu N)$&${\cal O}_{duNe}=(\overline{d}\gamma_\mu u)(\overline{N}\gamma^\mu e)$\\
\hline
LLRR&${\cal O}_{LN}=(\overline{L}\gamma_\mu L)(\overline{N}\gamma^\mu N)$&${\cal O}_{QN}=(\overline{Q}\gamma_\mu Q)(\overline{N}\gamma^\mu N)$\\
\hline
\multirow{2}{*}{\vtext{LRLR}}&${\cal O}_{LNLe}=(\overline{L} N)\epsilon (\overline{L}e)$&${\cal O}_{LNQd}=(\overline{L} N)\epsilon (\overline{Q} d)$\\
& \multicolumn{2}{c|}{${\cal O}_{LdQN}=(\overline{L}d)\epsilon (\overline{Q} N)$} \\
\hline
LRRL & \multicolumn{2}{c|}{${\cal O}_{QuNL}=(\overline{Q}u)(\overline{N}L)$} \\
\hline
\end{tabular}
\caption{\it Relevant four-fermion operators.}
\label{tab:fermionic}
\end{table}

At energies $E < M\equiv \mathrm{min}\left\lbrace M_{X_E}, M_{X_N}, 
M_{\varphi}\right\rbrace$, this model can be described by a local EFT built upon 
the SM fields and $N$, also known as $\nu$SMEFT. 
To leading order in the expansion in $E/M$, it is given by $\mathcal{L}_{SM+N}$ 
(with IR parameters) and a set of 
dimension-six operators:
\begin{equation}
 \mathcal{L}_{EFT} = \mathcal{L}_{SM+N}^{IR} + \frac{1}{\Lambda^2}\sum_i 
\alpha_i \mathcal{O}_i\,,
\end{equation}
with $\alpha_i$ being dimensionless couplings.
A basis of the operators $\mathcal{O}_i$~%
\footnote{We are not showing explicitly the CP counterparts 
of these operators, because we assume CP conservation.
However, they include: $i B_{\mu\nu} (\overline{N}\gamma^\mu\partial^\nu N)$ (the one without $i$, for both the normal field strength and for the dual, is redundant), $i\mathcal{O}_{NB}$, $i\mathcal{O}_{NW}$ (dipole operators with the dual are redundant), $i\mathcal{O}_{LN}^{1,2,3,4}$, $i\mathcal{O}_{LNH}$, $i\mathcal{O}_{HN}$ ($i\mathcal{O}_{NN}^2$ is redundant) and $i\mathcal{O}_{HNe}$. Note that $\partial^\nu\tilde{B}_{\mu\nu}(\overline{N}\gamma^\mu N)$  vanishes due to the Bianchi identity. 
On the four-fermion side, we would have $i\mathcal{O}_{duNe}$, $i\mathcal{O}_{LNLe}$, $i\mathcal{O}_{LNQd}$, $i\mathcal{O}_{LdQN}$ and $i\mathcal{O}_{QuNL}$.}, 
obtained with the help of \texttt{BasisGen}~\cite{Criado:2019ugp} 
(see Ref.~\cite{Fonseca:2019yya} for a similar code), 
is given in Tabs.~\ref{tab:bosonic} and \ref{tab:fermionic}. 
When evaluated on shell, the operators in grey 
can be removed from the action by suitable field redefinitions 
which, up to dimension-eight effects, can be implemented by using 
the equations of motion~\cite{Deans:1978wn,Politzer:1980me,Criado:2018sdb}. 
(Redundancies due to algebraic or Fierz identities or integration by parts have 
been removed.) Neglecting the small $m_N$ and the Yukawa couplings, the 
relevant equations of motion of $\mathcal{L}_{SM+N}$ read: 
\begin{align}
 i\slashed{\partial}N &= 0\,,\\
 i\slashed{D}L &=0\,, \\
 (D^2\tilde{H})^i &= \mu_H^2 \tilde{H}^i - \lambda_H (H^\dagger H)\tilde{H}^i\,,\\
 \partial^\nu B_{\nu\mu} &= -\frac{g'}{2} (i H^\dagger D_\mu H + \text{h.c.}) - g' Y^f \overline{f}\gamma_\mu f\,,
\end{align}
where $f$ runs over all SM$+N$ fermions. (The top Yukawa coupling is not negligible; however, its only impact would be the generation of four-fermion operators involving top quarks and $N$, for which there are no sensible searches.)
As a consequence, the following relations hold on shell for the operators in grey:
\begin{align}
 \mathcal{O}_{DN}^1 &= 0\,,\label{eq:odn1} \\
 \mathcal{O}_{DN}^2  &= - \mathcal{O}_{DN}^3\,, \\
   \mathcal{O}_{DN}^3 &= \frac{g'}{2}\mathcal{O}_{HN}  + g' Y^f\mathcal{O}_{fN}\,, \\
 \mathcal{O}_{LN}^1  &= \left(\mu_H^2 \overline{L}\tilde{H}N+\mathrm{h.c.}\right)-\lambda_H \mathcal{O}_{LNH} \,, \\
 \mathcal{O}_{LN}^2  &= - \mathcal{O}_{LN}^3 \,, \\
 \mathcal{O}_{LN}^3  &= \left(\frac{\mu_H^2}{2}\overline{L}\tilde{H}N +\mathrm{h.c.}\right) - 
 \frac{\lambda_H}{2}\mathcal{O}_{LNH} 
 +\frac{g'}{8}\mathcal{O}_{NB}-\frac{g}{8}\mathcal{O}_{NW}\,, \\
 \mathcal{O}_{LN}^4  &= 0\,, \\
 \mathcal{O}_{NN}^2  &= 0\,. \label{eq:onn2}
\end{align}
As a final remark, let us note that, in light of these equations, effective operators involving $N$ do not generate any purely SMEFT operators upon using the equations of motion.

\section{Matching}
\label{sec:matching}
%
Hereafter, we assume for simplicity $M_{X_E}=M_{X_N}=M_{\varphi}=M$. Also, we focus on the regime 
$g_X\sim g_L\sim \lambda_{\varphi H} \ll g_N$, and $g_N > 1$ 
(but $\lesssim 4\pi$ to stay in the perturbative regime).
This way, the mass and loop suppression in operators involving $N$ 
is compensated by the large $g_N$. 
On the other hand, purely SMEFT operators can be neglected. 

Our process of matching consists of equating one-light-particle-irreducible amplitudes computed in both the UV and the EFT at a scale $\mu=M$ in $\overline{MS}$ with space-time dimension $d = 4-2\epsilon$. Following the discussion above, we only compute those amplitudes involving $N$. 
Let us also note that, by virtue of Eq.~\eqref{eq:odn1}, the amplitude involving just two $N$ fields, to which only this operator contributes, does not need to be computed. Likewise, due to the absence of heavy particle couplings to $e$ in the UV Lagrangian, it can be trivially seen that $\alpha_{HNe} = 0$. 

\begin{figure}[t]
 \centering
 \includegraphics[width=\textwidth]{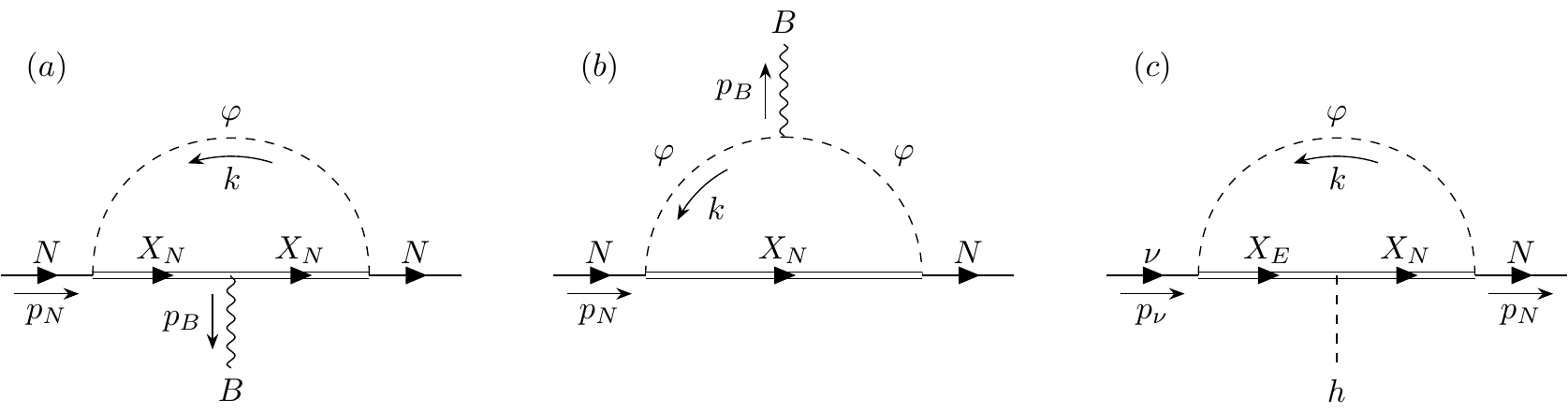}
 \caption{\it $(a)$ and $(b)$ Diagrams for the amplitude $\langle NNB\rangle$ in the UV, to which $\mathcal{O}_{DN}^2$ and $\mathcal{O}_{DN}^3$ contribute in the IR. $(c)$ Diagram for the amplitude $\langle\nu N h\rangle$ in the UV, to which $\mathcal{O}_{LN}^1$, $\mathcal{O}_{LN}^2$, $\mathcal{O}_{LN}^3$ and $\mathcal{O}_{LN}^4$ contribute in the IR.}
 \label{fig:NNB_and_vNH}
\end{figure}

The operators $\mathcal{O}_{DN}^{2}$ and $\mathcal{O}_{DN}^3$ can be matched by computing the amplitude 
given by the diagrams~%
\footnote{All Feynman diagrams in this article are produced 
with the \texttt{Ti\textit{k}Z-Feynman} package \cite{Ellis:2016jkw}.}
$(a)$ and $(b)$ in Fig.~\ref{fig:NNB_and_vNH}.
We use the momentum of the incoming $N$ 
and the momentum of the $B$, $p_N$ and $p_B$, respectively. 
In $\overline{MS}$ we drop terms proportional to 
$(1/\epsilon+\log{4\pi}-\gamma)$, 
where $\gamma$ is the Euler--Mascheroni constant.
The amplitudes in the UV and in the EFT to order $\mathcal{O}(p^2)$ read:
\begin{align}
 i\mathcal{M}_{UV} = \frac{ig^\prime g_N^2}{96\pi^2 M^2} 
 \overline{u}(p_N-p_B)P_L
 \bigg[& \gamma^\mu 
 \left(p_B^2 - p_B p_N + \slashed{p}_B \slashed{p}_N\right) \nonumber \\
 &- p_B^\mu \slashed{p}_B - p_B^\mu \slashed{p}_N + p_N^\mu \slashed{p}_B 
 \bigg] u(p_N)\epsilon^*_\mu(p_B)\,,
\end{align}
\begin{align}
 i\mathcal{M}_{EFT} = \frac{i}{\Lambda^2}\overline{u}(p_N-p_B)P_L \bigg[&\gamma^\mu \left(\alpha_{DN}^3p_B^2 - 2\alpha_{DN}^2p_B p_N +2\alpha_{DN}^2 \slashed{p}_B\slashed{p}_N\right) \nonumber \\
 &-\alpha_{DN}^3 p_B^\mu\slashed{p}_B - 2\alpha_{DN}^2p_B^\mu\slashed{p}_N + 2\alpha_{DN}^2p_N^\mu\slashed{p}_B\bigg] u(p_N)\epsilon^*_\mu(p_B)\,.
\end{align}
We provide all details about the computation of this and the forthcoming UV amplitudes in Appendix~\ref{app:UV}.

\begin{figure}[t]
 \centering
 \includegraphics[width=\textwidth]{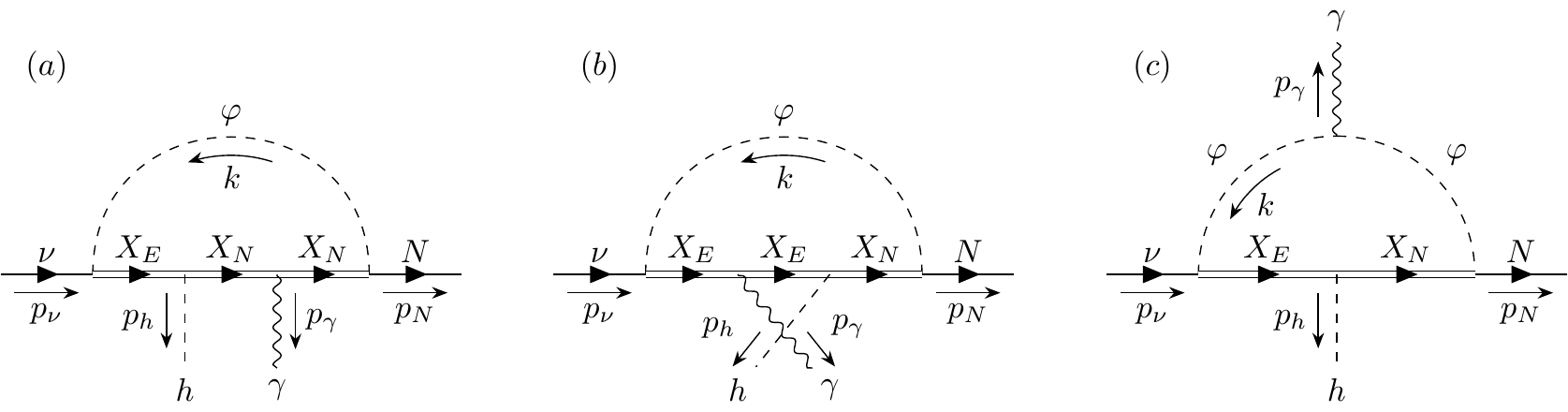}
 \caption{\it Diagrams for the amplitude $\langle\nu N h\gamma\rangle$ in the UV, to which $\mathcal{O}_{NB}$ and $\mathcal{O}_{NW}$ contribute in the IR.}
 \label{fig:vNAH}
\end{figure}

\begin{figure}[t]
 \centering
 \includegraphics[width=\textwidth]{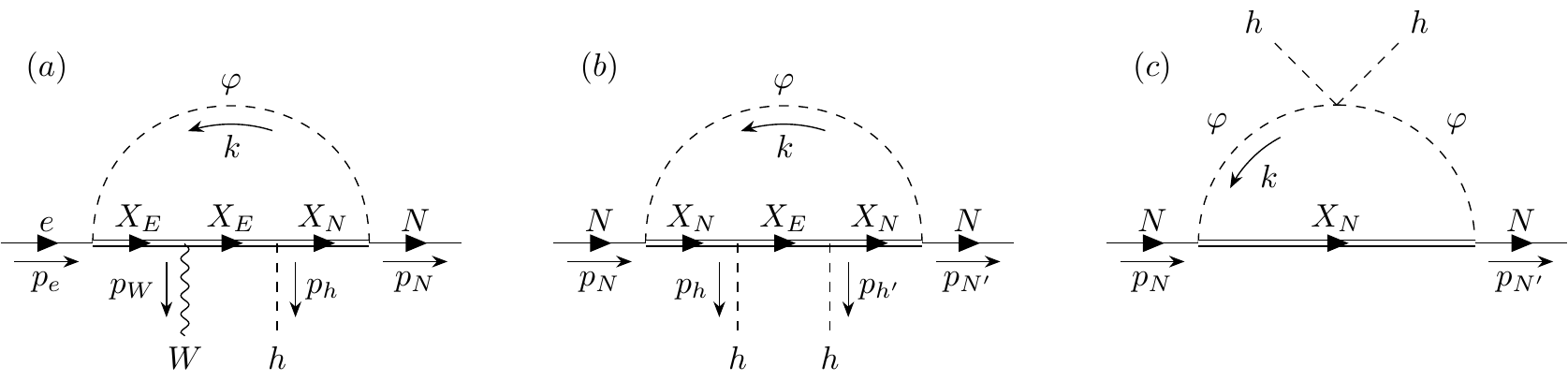}
 \caption{\it $(a)$ Diagram for the amplitude $\langle e N W h\rangle$ in the UV, to which $\mathcal{O}_{LN}^1$, $\mathcal{O}_{LN}^2$, $\mathcal{O}_{LN}^3$ and $\mathcal{O}_{NW}$ contribute in the IR. $(b)$ and $(c)$ Diagrams for the amplitude $\langle NN h h\rangle$ in the UV, to which $\mathcal{O}_{NN}^2$ contributes in the IR. (Note that, despite not explicitly shown, diagram~$(b)$ but with the two Higgses exchanged is also present.)}
 \label{fig:eNWH_and_NNHH}
\end{figure}

The operators $\mathcal{O}_{LN}^1$, $\mathcal{O}_{LN}^2$, $\mathcal{O}_{LN}^3$ and $\mathcal{O}_{LN}^4$, as well as $Y_N$ in the IR, can be matched by computing the amplitude represented by the diagram $(c)$ in Fig.~\ref{fig:NNB_and_vNH}. We take $p_\nu$ and $p_N$ as independent momenta. 
To order $\mathcal{O}(p^2)$ we have: 
\begin{equation}
 i\mathcal{M}_{UV} = \frac{i g_N g_X g_L}{96\sqrt{2}\pi^2M^2}\overline{u}(p_N) P_L 
 \bigg[ 6M^2\left(1-\log{\frac{\mu^2}{M^2}}\right) - p_\nu^2 - p_N^2 + p_{\nu}p_N 
 + \slashed{p}_N\slashed{p}_\nu \bigg] u(p_\nu)\,,
 \label{eq:vHN}
\end{equation}
\begin{align}
 i\mathcal{M}_{EFT} = \frac{i}{\sqrt{2}\Lambda^2}\overline{u}(p_N) P_L\bigg[&-Y_N\Lambda^2 - \alpha_{LN}^1 p_\nu^2 + \left(\alpha_{LN}^2-\alpha_{LN}^1-\alpha_{LN}^4\right)p_N^2 \nonumber \\
 &+\left(2\alpha_{LN}^1-\alpha_{LN}^2+\alpha_{LN}^3\right) p_\nu p_N 
-\alpha_{LN}^3\slashed{p}_N\slashed{p}_\nu\bigg] u(p_\nu)\,.\label{eq:vHNeft}
\end{align}
The other two operators involving a single Higgs field are $\mathcal{O}_{NB}$ and $\mathcal{O}_{NW}$. They can be matched by computing the amplitude represented by the diagrams in Fig.~\ref{fig:vNAH}.
Taking $p_\gamma$, $p_h$ and $p_N$ as independent momenta, 
the results in the UV and in the EFT up to order $\mathcal{O}(p)$ read: 
\begin{equation}
 i\mathcal{M}_{UV} = 
 \frac{i g_L g_X g_N e}{96\sqrt{2}\pi^2 M^2}\overline{u}(p_N)P_L 
 \left[\gamma^\mu\slashed{p}_\gamma - p_\gamma^\mu\right] u(p_\nu) \epsilon_\mu^*(p_\gamma)\,,
\end{equation}
\begin{equation}
 i\mathcal{M}_{EFT} = \frac{\sqrt{2}i}{\Lambda^2}\left(c_W \alpha_{NB} + s_W \alpha_{NW}\right)
 \overline{u}(p_N)P_L
 \left[\gamma^\mu\slashed{p}_\gamma-p_\gamma^\mu\right] u(p_\nu) \epsilon_\mu^*(p_\gamma)\,.
\end{equation}
Here $c_W \equiv \cos \theta_W$ and $s_W \equiv \sin \theta_W$, 
with $\theta_W$ being the weak mixing angle.
(Let us emphasise that in our convention, $W^3_\mu = c_W Z_\mu + s_W A_\mu$, $B_\mu = c_W A_\mu - s_W Z_\mu$.)
This amplitude was also computed previously in Ref.~\cite{Butterworth:2019iff} (see appendix therein).
Still, one more amplitude needs to be computed in order to completely fix the Wilson coefficients of the one-Higgs operators. 
We choose that represented by the diagram $(a)$ in Fig.~\ref{fig:eNWH_and_NNHH}.
Taking $p_h$, $p_W$ and $p_N$ as independent momenta, we have to order $\mathcal{O}(p)$:
\begin{equation}
 i\mathcal{M}_{UV} = \frac{i g_N g_X g_L g}{192 \pi^2 M^2} 
 \overline{u}(p_N) P_L \bigg[ p_N^\mu - 2p_h^\mu - p_W^\mu - \gamma^\mu 
\slashed{p}_N \bigg] u(p_e) \epsilon_\mu^\ast(p_W)\,,\label{eq:eNWh}
\end{equation}
\begin{align}
 i\mathcal{M}_{EFT} = \frac{i g}{2\Lambda^2}\overline{u}(p_N) P_L\bigg[&- \left(\alpha_{LN}^2+\alpha_{LN}^3 \right) p_N^\mu -2\alpha_{LN}^1p_h^\mu -\left(4\frac{\alpha_{NW}}{g}+\alpha_{LN}^1\right)p_W^\mu \nonumber\\
 &+\alpha_{LN}^3\gamma^\mu\slashed{p}_N + 
4\frac{\alpha_{NW}}{g}\gamma^\mu\slashed{p}_W \bigg] u(p_e) 
\epsilon_\mu^*(p_W)\,.\label{eq:eNWheft}
\end{align}

\begin{figure}[t]
 \centering
 \includegraphics[width=\textwidth]{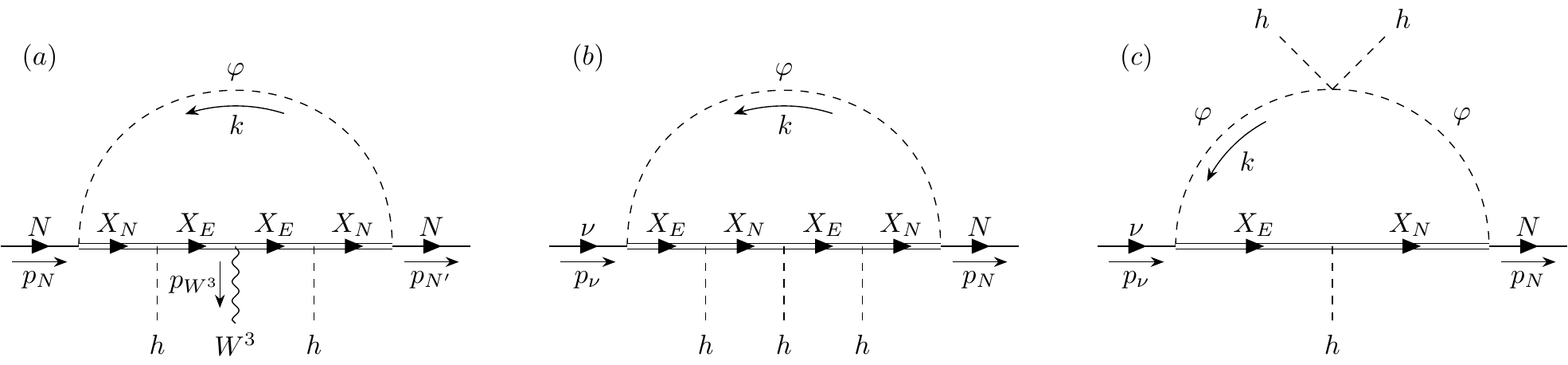}
 \caption{\it $(a)$ Diagram for the amplitude $\langle NNW^3 h h\rangle$ in the UV, to which the operator $\mathcal{O}_{HN}$ contributes in the IR. $(b)$ and $(c)$ Diagrams for the amplitude $\langle\nu N hhh\rangle$ in the UV, to which $\mathcal{O}_{LNH}$ contributes in the IR. (Note that, despite not explicitly shown, all these diagrams but with the corresponding Higgs legs exchanged are also present.)}
 \label{fig:NNHHW3_and_vNHHH}
\end{figure}

The operator $\mathcal{O}_{NN}^2$ can be matched by computing the amplitude represented by the diagrams~$(b)$ and $(c)$ 
in Fig.~\ref{fig:eNWH_and_NNHH}, 
while the operator $\mathcal{O}_{HN}$ by computing 
the diagram $(a)$ in Fig.~\ref{fig:NNHHW3_and_vNHHH}.
(It might seem that $\mathcal{O}_{HN}$ also contributes to the former amplitude; however, only its CP counterpart $i\mathcal{O}_{HN}$, which we do not need to consider, does it.) 
The first one reads, to order $\mathcal{O}(p)$:
\begin{equation}
 i\mathcal{M}_{UV} = - \frac{ig_N^2 \lambda_{\varphi H}}{96\pi^2M^2} 
 \overline{u}(p_{N'}) P_L \left[\slashed{p}_N + \slashed{p}_{N'}\right] u(p_N)\,,
\end{equation}
\begin{equation}
 i\mathcal{M}_{EFT} = \frac{i}{\Lambda^2}\alpha_{NN}^2 \overline{u}(p_{N'}) P_L \left[\slashed{p}_{N} + \slashed{p}_{N'}\right] u(p_N)\,.
\end{equation}
In the UV and EFT to zero momentum, the second aforementioned amplitude reads:
\begin{equation}
 i\mathcal{M}_{UV} = \frac{i g g_N^2 g_{X}^2}{96\pi^2 M^2} \overline{u}(p_{N'}) P_L \gamma^\mu u(p_N) \epsilon_\mu^*(p_{W^3})\,,
\end{equation}
\begin{equation}
 i\mathcal{M}_{EFT} = -\frac{i\alpha_{HN} g}{\Lambda^2} \overline{u}(p_{N'}) P_L \gamma^\mu u(p_N) \epsilon_\mu^*(p_{W^3})\,.
\end{equation}
The operator $\mathcal{O}_{LNH}$ can be matched by computing the amplitude depicted by the diagrams~$(b)$ and $(c)$ 
in Fig.~\ref{fig:NNHHW3_and_vNHHH}. 
To zero momentum, the amplitudes in the UV and in the EFT are given by~\cite{Butterworth:2019iff}: 
\begin{equation}
 i\mathcal{M}_{UV} = \frac{i g_N g_X g_L}{32\sqrt{2}\pi^2M^2} 
 \left(\lambda_{\varphi H} - g_X^2\right) 
 \overline{u}(p_N) P_L  u(p_\nu)\,,
\end{equation}
\begin{equation}
 i\mathcal{M}_{EFT} = \frac{3i\alpha_{LNH}}{\sqrt{2}\Lambda^2}\overline{u}(p_N)P_L u(p_\nu)\,.
\end{equation}

\begin{figure}[t]
 \centering
 \includegraphics[width=0.66\textwidth]{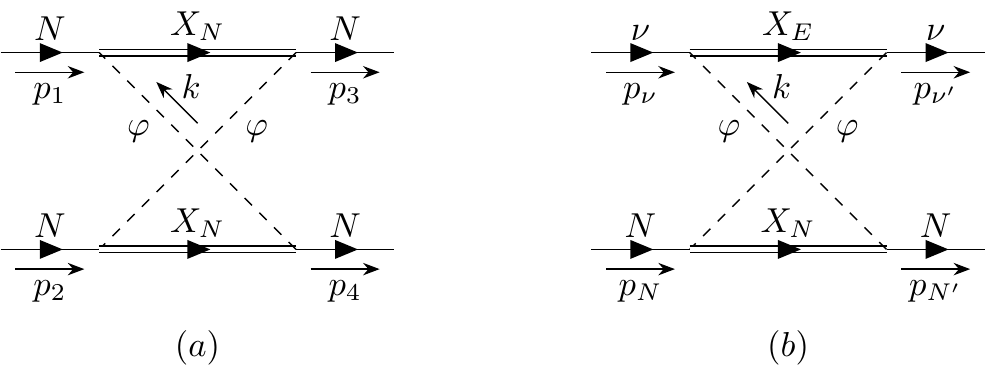}
 \caption{\it $(a)$ Diagram for the amplitude $\langle NNNN\rangle$ in the UV, to which $\mathcal{O}_{NN}$ contributes in the IR. (Note that, despite not explicitly shown, this diagram but with the two outgoing $N$s exchanged also exists.) $(b)$ Diagram for the amplitude $\langle \nu \nu N N\rangle$ in the UV, to which $\mathcal{O}_{LN}$ contributes in the IR.}
 \label{fig:NNNN_and_NNvv}
\end{figure}

On the side of four-fermion operators, the only such non-vanishing interactions (before using the equations of motion) are $\mathcal{O}_{NN}$ and $\mathcal{O}_{LN}$. They can be matched by computing the amplitudes depicted by the diagrams~$(a)$ and $(b)$ 
in Fig.~\ref{fig:NNNN_and_NNvv}, respectively. 
The UV and EFT expressions for each amplitude 
to zero momentum read, respectively:
\begin{equation}
 i\mathcal{M}_{UV} = - \frac{i g_N^4}{96 \pi^2 M^2}
 \left[\overline{u}(p_3)\gamma^\mu P_R u(p_1)\right]
 \left[\overline{u}(p_4)\gamma_\mu P_R u(p_2)\right],
\end{equation}
\begin{equation}
 i\mathcal{M}_{EFT} = \frac{4i\alpha_{NN}}{\Lambda^2}
 \left[\overline{u}(p_3)\gamma^\mu P_R u(p_1)\right]
 \left[\overline{u}(p_4)\gamma_\mu P_R u(p_2)\right],
\end{equation}
and
\begin{equation}
 i\mathcal{M}_{UV} = -\frac{ig_N^2 g_L^2}{192\pi^2 M^2} 
 \left[\overline{u}(p_{\nu'})\gamma^\mu P_L u(p_\nu)\right]
 \left[\overline{u}(p_{N'})\gamma_\mu P_R u(p_N)\right],
\end{equation}
\begin{equation}
 i\mathcal{M}_{EFT} = \frac{i\alpha_{LN}}{\Lambda^2}
 \left[\overline{u}(p_{\nu'})\gamma^\mu P_L u(p_\nu)\right]
 \left[\overline{u}(p_{N'})\gamma_\mu P_R u(p_N)\right].
\end{equation}

By equating all UV amplitudes to their IR counterparts, 
we end up with 23 equations (including  $\alpha_{HNe}=0$) 
for 15 unknowns; the redundancies reflect the gauge 
symmetries~\footnote{Note, for example, that the 
contributions of $\mathcal{O}_{LN}^1$ to the amplitudes $\langle\nu N h 
\rangle$ and $\langle e N W h\rangle$ are correlated, because both the 
$\mathcal{O}(p^2)$ Higgs piece as well as the $WH$ interaction come from $D^2\tilde{H}$. Thus, when matching \textit{e.g.} the $p_\nu^2$ part of Eqs.~\eqref{eq:vHN} 
and \eqref{eq:vHNeft} one gets $\alpha_{LN}^1 = (g_N g_X g_L)/(96\pi^2)\times 
(\Lambda/M)^2$; exactly the same as matching the $p_h^\mu$ piece of 
Eqs.~\eqref{eq:eNWh} and \eqref{eq:eNWheft}.}.
Neglecting the running from the scale $\mu=M$ to $\mu=v$, the following identities hold off shell at the EW scale (other Wilson coefficients vanish): 
\begin{gather}
 Y_N^{IR} = Y_N^{UV} - \frac{g_L g_X g_N}{16\pi^2}\,, \\
 \begin{align}
 \frac{\alpha_{DN}^2}{\Lambda^2} &= \frac{e g_N^2}{192 \pi^2 c_W M^2}\,, \Label{} &
 \frac{\alpha_{DN}^3}{\Lambda^2} &= \frac{e g_N^2}{96 \pi^2 c_W M^2}\,, \\
 \frac{\alpha_{NB}}{\Lambda^2} &= \frac{e g_L g_X g_N}{192 \pi^2 c_W M^2}\,, \Label{} &
 \frac{\alpha_{LN}^1}{\Lambda^2} &= \frac{g_L g_X g_N}{96\pi^2M^2}\,, \\
 \frac{\alpha_{LN}^3}{\Lambda^2} &= -\frac{g_L g_X g_N}{96 \pi^2 M^2}\,, \Label{} &
 \frac{\alpha_{HN}}{\Lambda^2} &= -\frac{g_X^2 g_N^2}{96\pi^2 M^2}\,, \\
 \frac{\alpha_{NN}^2}{\Lambda^2} &= -\frac{\lambda_{\varphi H} g_N^2}{96\pi^2 M^2}\,, \Label{} &
 \frac{\alpha_{LNH}}{\Lambda^2} &= \frac{g_L g_X g_N}{96\pi^2 M^2} (\lambda_{\varphi H}-g_X^2)\,, \\
 \frac{\alpha_{NN}}{\Lambda^2} &= -\frac{g_N^4}{384\pi^2 M^2}\,, \Label{} &
 \frac{\alpha_{LN}}{\Lambda^2} &= -\frac{g_L^2 g_N^2}{192\pi^2 M^2}\,, 
 \end{align}
\end{gather}
where $e=\sqrt{4\pi\alpha}$, and $\alpha\approx 1/137$ stands for the electromagnetic fine-structure constant. Finally,
upon using the equations of motion, 
Eqs.~\eqref{eq:odn1}--\eqref{eq:onn2}, 
the following relations hold on shell (other Wilson coefficients vanish): 
\begin{gather}
 Y_N^{IR} = Y_N^{UV} - \frac{g_L g_X g_N}{16 \pi^2} \left(1+ \frac{m_h^2}{24 M^2}\right)\,, \label{eq:yn} \\
 \begin{align*}
 \frac{\alpha_{NB}}{\Lambda^2} &= \frac{e g_L g_X g_N}{256 \pi^2 c_W M^2}\,, \Label{} &
 \frac{\alpha_{NW}}{\Lambda^2} &= \frac{e g_L g_X g_N}{768 \pi^2 s_W M^2}\,, \Label{} \\ 
 \frac{\alpha_{HN}}{\Lambda^2} &= \frac{g_N^2 (e^2-4c_W^2 g_X^2)}{384 \pi^2 c_W^2 M^2}\,, \Label{} &
 \frac{\alpha_{LNH}}{\Lambda^2} &= -\frac{g_L g_X g_N}{192\pi^2 M^2}\left[\frac{m_h^2}{v^2} + 2 (g_X^2-\lambda_{\varphi H})\right]\,, \Label{} \\
 \frac{\alpha_{LN}}{\Lambda^2} &= -\frac{g_N^2 (e^2+2c_W^2g_L^2)}{384 \pi^2 c_W^2 M^2}\,, \Label{} &
 \frac{\alpha_{eN}}{\Lambda^2} &= -\frac{e^2 g_N^2}{192 \pi^2 c_W^2 M^2}\,, \Label{} \\
 \frac{\alpha_{NN}}{\Lambda^2} &= -\frac{g_N^4}{384\pi^2 M^2}\,, \Label{} &
 \frac{\alpha_{QN}}{\Lambda^2} &= \frac{e^2 g_N^2}{1152 \pi^2 c_W^2 M^2}\,, \Label{} \\
 \frac{\alpha_{uN}}{\Lambda^2} &= \frac{e^2 g_N^2}{288 \pi^2 c_W^2 M^2}\,, \Label{} &
 \frac{\alpha_{dN}}{\Lambda^2} &= -\frac{e^2 g_N^2}{576 \pi^2 c_W^2 M^2}\,. \Label{eq:dN}
\end{align*}
\end{gather}
For convenience, let us also define 
$\mathcal{O}_{NA} = c_W\mathcal{O}_{NB} + s_W\mathcal{O}_{NW}$ 
and $ \mathcal{O}_{NZ} =  c_W\mathcal{O}_{NW} -s_W\mathcal{O}_{NB}$. 
For the coefficients of these operators we obtain:
\begin{align}
 \frac{\alpha_{NA}}{\Lambda^2} &= \frac{e g_L g_X g_N}{192\pi^2 M^2}\,, \Label{} &
\frac{\alpha_{NZ}}{\Lambda^2} &= \frac{e g_L g_X g_N (1-4s_W^2)}{768 \pi^2 s_W c_W M^2}\,.
\end{align}

Finally, let us emphasise that the coupling constants 
of renormalisable operators in the EFT can 
also be written in terms of the UV couplings; $g_{IR} = 
g_{UV} + g'_{UV}/(16\pi^2)$. However, one can always reabsorb the one-loop 
corrections by redefining $g_{UV}\to g_{UV}-g'_{UV}/(16\pi^2)$. This 
redefinition would propagate to all $\alpha$ couplings, but the impact would be 
formally of two-loop order. Still, for the sake of 
completeness, we have computed all renormalisable EFT terms in Appendix 
\ref{app:dim4}.

\section{Sterile neutrino phenomenology}
\label{sec:pheno}
%
In the process of matching we have neglected $m_N$. The only effect of $m_N\ne 0$ would appear in the dimension-five operator $\overline{N^c} N H^\dagger H$ suppressed not only by the loop factor but also by $m_N/M$, namely by $\sim 10^{-3}$ if $m_N\sim 1$ GeV and $M\sim 1$ TeV. (The operator $\overline{N^c}\sigma^{\mu\nu} N B_{\mu\nu}$ vanishes in this case because $N$ is Majorana.) However, $m_N\neq0$ has a huge impact on the phenomenology of $N$, because it allows the latter to decay into $\nu\gamma$. The corresponding decay width must be computed after EWSB, namely in the $\nu$LEFT, obtained first by matching the $\nu$SMEFT at the EW scale and after running down to $\sim m_N$. The full list of $\nu$LEFT operators involving $N$ is given in Tab.~\ref{tab:fermionic2} in Appendix~\ref{app:left}. The operator that triggers the decay of $N$ is $\mathcal{O}_{N\gamma}$. For completeness, though, we provide tree-level matching of all $\nu$SMEFT operators to all $\nu$LEFT ones in Eqs.~\eqref{eq:match1}--\eqref{eq:match2}. The one-loop running of all $\nu$LEFT operators generated in our setup, including $\mathcal{O}_{N\gamma}$, is also given in the same appendix. Altogether, we have:
\begin{equation}
 \Gamma(N\to\nu\gamma) \approx 
 \frac{m_N^3 \alpha_{N\gamma}^2(v)}{2\pi v^2}
 \left(1 - \frac{5e^2}{9\pi^2} \log\frac{v}{m_N} \right)^2\,.
 \label{eq:GammaNRG}
\end{equation}
For simplicity, let us fix $Y_N^{UV}$ such that $Y_N^{IR} = \alpha_{LNH} v^2/(2\Lambda^2)$; see Eq.~\eqref{eq:yn}. 
This way, the mixing between $N$ and $\nu$ vanishes strictly. 
(Note that the sole important effect of this mixing would be inducing a Majorana neutrino dipole moment for $\nu$; this vanishes however in our case due to lepton flavour conservation~\cite{Butterworth:2019iff}.) 

Different experiments constrain the parameter space under study. 
The relevant observables can be all computed directly in the $\nu$SMEFT. (In quoting the following bounds we have set $\Lambda = 1$ TeV.) We have first $\mathcal{B}(Z\to\nu\nu\gamma\gamma)$. Experimentally it is bounded to be $<3.1\times 10^{-6}$~\cite{Tanabashi:2018oca}. In our context this branching ratio is given by (note that for $m_N \sim 1$~GeV, $\mathcal{B}(N\to\nu\gamma)\approx 1$~\cite{Duarte:2015iba}):
\begin{align}
 \mathcal{B}(Z\to\nu\nu\gamma\gamma) \approx\mathcal{B}(Z\to NN) 
 &\approx \frac{1}{\Gamma_Z^{SM}}\frac{m_Z^3 v^2}{24\pi\Lambda^4}\alpha_{HN}^2 \nonumber\\
 &\approx \, 2.7\times 10^{-10} g_N^4 
 \left(0.029-g_X^2\right)^2\frac{\text{TeV}^4}{M^4}\,,
\end{align}
with $\Gamma_Z^{SM}\approx 2.5$ GeV.
This bound implies in turn a limit on $\alpha_{HN} < 0.11$, which is ultimately the most stringent constrain on $(M, g_N)$. 
Other subleading constraints include: \textit{(i)} $\mathcal{B}(Z\to\nu\nu\gamma)$, experimentally bounded to be $ < 3.2\times 10^{-6}$~\cite{Acciarri:1997im}, which implies $\alpha_{NZ} < 0.081$;
\textit{(ii)} the measurement of the total $W$ boson width, $\Gamma_W^\text{total} = 2.085\pm 0.042$ GeV, which however does not constrain $\alpha_{HNe}$ more than a theoretical perturbativity bound implying $\alpha_{HNe}<4\pi$; \textit{(iii)} the bound on $\alpha_{NA} < 0.88$~\cite{Butterworth:2019iff} as obtained from LHC searches for events with one photon and missing energy~\cite{Sirunyan:2018dsf}. 
(Bounds on $\alpha_{NA}$ obtained from the study of differential Drell-Yan distributions at the LHC, mediated by both neutral and charged currents, using \texttt{Contur}~\cite{Butterworth:2016sqg} are weaker~\cite{Butterworth:2019iff}.) 
The bound on $\alpha_{NA}$ can be improved to 
$\alpha_{NA} < 0.36$ by searches for 
$h \to \gamma\gamma + E_T^\text{miss}$, 
as proposed in Ref.~\cite{Butterworth:2019iff}. 
This value, however, still leads to a very weak constrain on $(M,g_N)$.

\begin{figure}[t]
 \centering
 \includegraphics[width=0.48\textwidth]{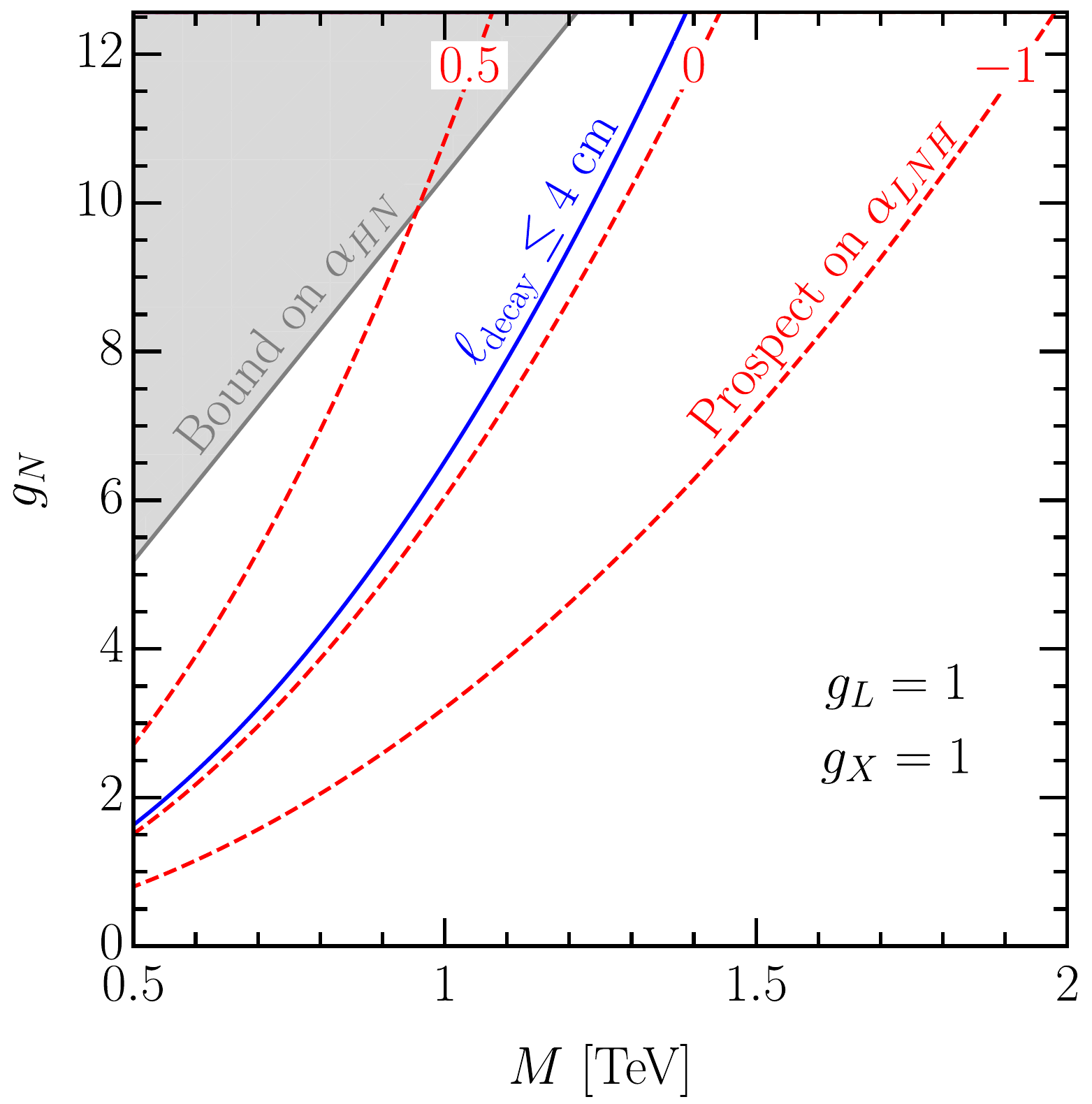}
 \hspace{0.02\textwidth}
 \includegraphics[width=0.48\textwidth]{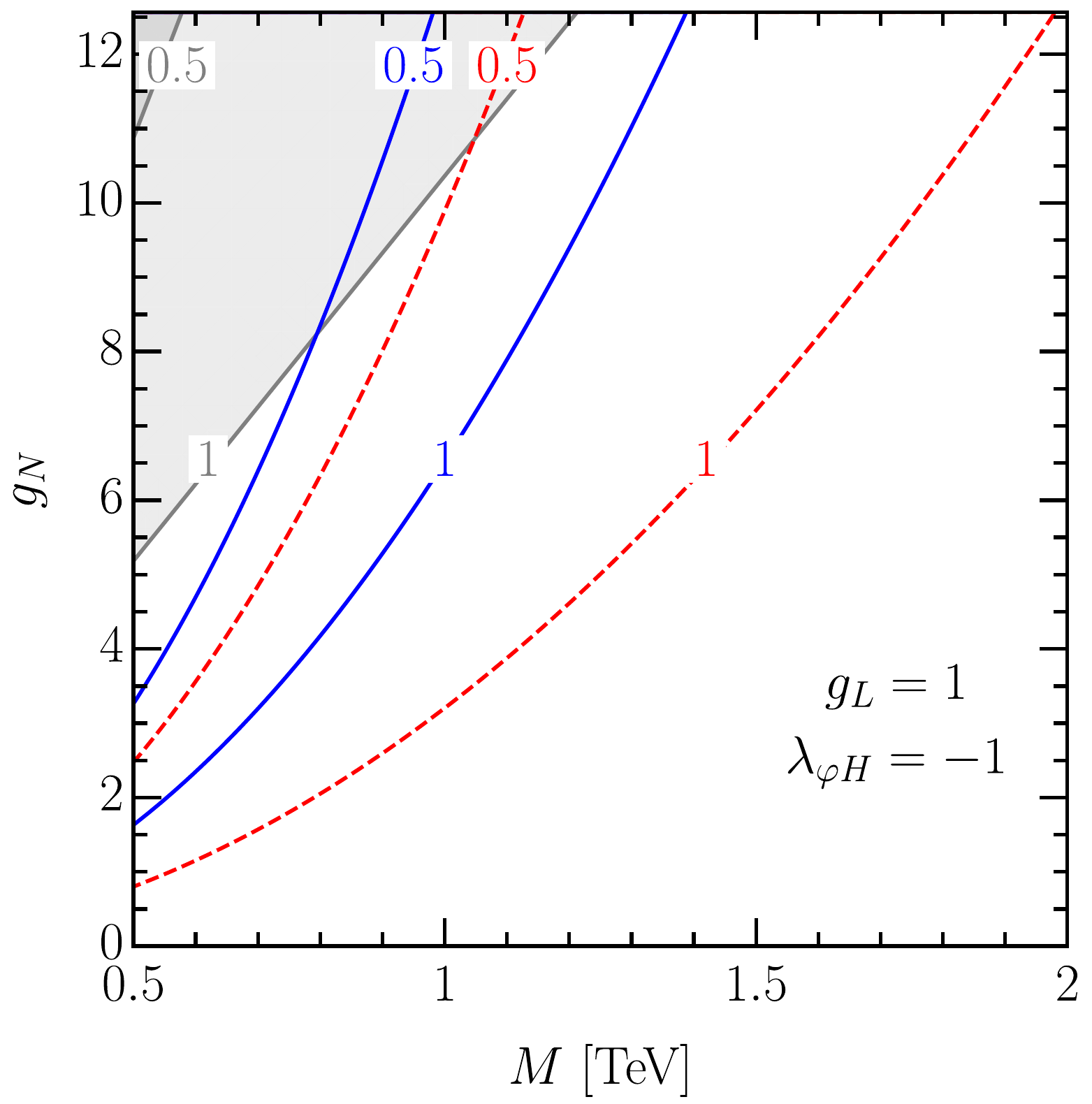}
 \caption{\it Constraints in the plane $(M,g_N)$ derived from 
 the bounds on the EFT coefficients summarised in the text.
 In the region above the blue line, $N$ with $m_N = 1$~GeV
 decays within $4$\,cm. The actual bound on $\alpha_{HN}$ 
 from $Z \to \nu\nu\gamma\gamma$
 and the prospective bound on $\alpha_{LNH}$ 
 from the $h \to \gamma + E_T^\text{miss}$ 
 analysis proposed in Ref.~\cite{Butterworth:2019iff}
 have been translated to the constraints in the plane $(M, g_N)$
 assuming $g_X = 1$ and 
 $\lambda_{\varphi H} = -1$, $0$ and $0.5$ (left) as well as 
 $\lambda_{\varphi H} = -1$ and $g_X = 0.5$ and $1$ (right).}
 \label{fig:constraints}
\end{figure}

Four-fermion interactions could be bounded at the LHC in searches for $pp \to \ell\gamma+ E_T^\text{miss}$. However, we are not aware of any such search; a preliminary phenomenological study has been provided in Ref.~\cite{Duarte:2016caz}.
Interestingly though, it has been shown that searches for Higgs decaying to a single photon and missing energy could test 
$\mathcal{B}(h\to \nu N) > 1.2\times 10^{-4}$~\cite{Butterworth:2019iff}.
Noticing that
\begin{align}
 \mathcal{B}(h\to \nu N) &\approx \frac{1}{\Gamma_h^{SM}} 
 \frac{m_h v^4}{16\pi\Lambda^4} \alpha_{LNH}^2 \nonumber \\
 &\approx 2.5\times 10^{-6} \left(g_L g_N g_X\right)^2 (0.13 + g_X^2-\lambda_{\varphi H})^2\frac{\text{TeV}^4}{M^4}\,,
\end{align}
with $\Gamma_h^{SM}\approx 4$ MeV, 
the corresponding limit on $\alpha_{LNH}$ reads 
$\alpha_{LNH} < 7.3\times 10^{-3}$.
We show in Fig.~\ref{fig:constraints} that, 
when translated to the plane $(M, g_N)$, 
this signal overcomes often the constraint from $\alpha_{HN}$.

If lepton number is exactly conserved, 
\textit{i.e.} in particular, $m_N = 0$, 
$N$ is just the RH component of the SM neutrino, 
which would be Dirac. 
In this case, the very stringent bounds on the neutrino
dipole moment~\cite{Canas:2015yoa} can be only satisfied if $g_X\approx 0$ (or $g_L \approx 0$). 
Accordingly, only the operator $\mathcal{O}_{HN}$ and
the four-fermions $\mathcal{O}_{LN}$, $\mathcal{O}_{eN}$,
$\mathcal{O}_{QN}$, $\mathcal{O}_{uN}$
and $\mathcal{O}_{dN}$ as well as $\mathcal{O}_{NN}$ could survive,
see Eqs.~\eqref{eq:yn}--\eqref{eq:dN}. 
The former enhances the $Z$ decay into invisible, 
but the corresponding limit on $(M, g_N)$ is very weak. 
Likewise, the bounds on the four-fermion
operators involving quarks and charged leptons are of order 
$\alpha/\Lambda^2 \lesssim 1$~TeV$^{-2}$~\cite{Alcaide:2019pnf}, 
and therefore they are not stringent in this setup 
in which all operators arise at loop level, 
and hence the effective scale $\Lambda$ is rather 
$\Lambda\sim 4\pi M$.

Finally, we are not aware of any significant bound on $\mathcal{O}_{NN}$.
As things stand, this scenario is very much unconstrained 
in light of current data, even for $M\sim $ few hundreds GeV. 
(In this respect, let us also emphasise that direct LHC searches 
for singly charged scalars and vector-like leptons, 
which are present in our UV model, do not constrain this range of masses~\cite{Redi:2013pga,Cao:2017ffm,Aaboud:2018jiw,Alcaide:2019kdr}.)

In the lepton number conserving case, 
if instead of a single $N$ we have three copies 
(in which case all neutrinos would be Dirac), with \textit{a priori} 
flavour-generic couplings, the off-diagonal dipole operators would induce 
decays of the SM neutrinos, $\nu_j\to\nu_i\gamma$.
For massless neutrino in the final state and neglecting running, 
the corresponding decay width reads
\begin{equation}
\Gamma(\nu_j\to\nu_i\gamma) = 
\frac{m_j^3 v^2}{8\pi\Lambda^4} \alpha_{NA}^2 = 
\frac{m_j^3 v^2}{8\pi M^4} \left(\frac{e g_L g_X g_N}{192\pi^2}\right)^2,
\end{equation}
where $m_j$ is the mass of $\nu_j$.
The lower limit on the neutrino lifetime is $\tau \gtrsim 10^{20}$~s~\cite{Tanabashi:2018oca,Aalberts:2018obr}. 
Assuming $m_j \approx 0.1$~eV and $\mathcal{O}(1)$ couplings, 
we obtain $M \gtrsim 0.3$~TeV,
and up to $\sqrt{4\pi}$ times 
more stringent constraint if $g_N$ is significantly larger than $1$, 
as we have been assuming in this work.
Still, as in the previous case, this bound can be avoided if 
\textit{e.g.} $g_X\ll 1$.

\section{Conclusions}
\label{sec:conclusions}
%
In this paper, we have considered a very simple extension of the SM 
involving a light RH neutrino $N$ (which can be well the RH part of 
any of the SM neutrinos if they are Dirac, or a new Majorana neutrino),
two heavy fermionic fields and one heavy scalar field, 
all of them colourless. 
In the IR, this theory can be described by the $\nu$SMEFT. 
We have shown that, if $N$ is Majorana, new Higgs decays 
not yet studied experimentally at the LHC can test this model better 
than other studies already performed at low-energy facilities; 
most importantly, searches for $Z\to\nu\nu\gamma\gamma$.

We note that, although this observation is relatively straightforward 
in the generic EFT, because constrained operators can be set to zero 
independently of those triggering the signal of interest, 
this is highly non-trivial in the EFT obtained in this model, 
in which all Wilson coefficients depend on solely four 
arbitrary couplings.
The fact that the signal of interest is not in conflict 
with the present constraints in such a simple UV completion 
of the $\nu$SMEFT, strengthens the motivation 
for novel searches in the Higgs sector.

We have provided a complete calculation of one-loop matching 
in the diagrammatic approach, obtained upon computing 
the same one-light-particle-irreducible off-shell amplitudes 
in the UV and in the IR.
This complements the very few examples of one-loop
matching in the literature, and it is expected that our results 
will allow faster progress in the automation of tools
in this respect~\cite{delAguila:2016zcb,Criado:2017khh}. 
As a byproduct of this work, we have also obtained a complete 
(off-shell) basis in the $\nu$SMEFT, the tree-level matching of the 
$\nu$SMEFT onto the low-energy version 
(in which the top quark, the Higgs and the $W$ and $Z$ bosons 
are integrated out), that we dubbed $\nu$LEFT; as well as
some one-loop anomalous dimensions in the latter EFT.
We leave the computation of the full RGE anomalous dimension matrix in the $\nu$LEFT and in the $\nu$SMEFT for future work.

\section*{Acknowledgements}
%
We are grateful to Jose Santiago for helpful discussions.
MC is supported by the Spanish MINECO under the Juan de la Cierva programme.

\appendix
%
%
\section{Mathematical tools}
\label{app:tools}
%
We have used the following master integrals:
\begin{align}
 \int \frac{\mathrm{d}^d k}{(2\pi)^d} \frac{1}{(k^2 - M^2)^n} &= \frac{(-1)^n 
i}{(4\pi)^{d/2}} \frac{\Gamma(n-d/2)}{\Gamma(n)} \frac{1}{M^{2n-d}} 
= A_n\,,\\\nonumber\\
\int \frac{\mathrm{d}^d k}{(2\pi)^d} \frac{k^\mu k^\nu}{(k^2 - M^2)^n} &= 
\underbrace{\frac{1}{2}\frac{(-1)^{n-1} i}{(4\pi)^{d/2}} 
\frac{\Gamma(n-d/2-1)}{\Gamma(n)} \frac{1}{M^{2n-d-2}}}_{B_n} 
g^{\mu\nu}\,,\\\nonumber
\int \frac{\mathrm{d}^d k}{(2\pi)^d} \frac{k^\mu k^\nu k^\rho k^\sigma}{(k^2 - M^2)^n} 
&= \overbrace{\frac{1}{4}\frac{(-1)^n i}{(4\pi)^{d/2}} 
\frac{\Gamma(n-d/2-2)}{\Gamma(n)}\frac{1}{M^{2n-d-4}}}^{C_n}\,\\
&\phantom{{}={}}\times\left(g^{\mu\nu}g^{\rho\sigma} + g^{\mu\rho} g^{\nu\sigma} + 
g^{\mu\sigma}g^{\nu\rho}\right)\,.
\end{align}
Here $d$ is the space-time dimension.
For expansion in an external momentum $p$ we have: 
\begin{equation}
 \frac{1}{(k+p)^2-M^2} = \frac{1}{k^2-M^2}\bigg[1 - 
 \frac{2kp+p^2}{k^2-M^2} + \frac{4 (kp)^2}{(k^2-M^2)^2} \bigg] + 
 \mathcal{O}(p^3)\,.
\end{equation}
Finally, we have also made use of the following algebraic identities:
\begin{equation}
 \epsilon^{\mu\sigma\rho\nu} p_{1\rho} p_{2\nu} \gamma_\sigma \gamma_5 = i \left(\gamma^\mu\slashed{p}_2\slashed{p}_1 + p_1^\mu\slashed{p}_2 - p_2^\mu\slashed{p}_1-\gamma^\mu p_1 p_2\right)\,,
\end{equation}
\begin{equation}
 [D_\mu, D_\nu] = -ig' YB_{\mu\nu} - igT^I W^{I}_{\mu\nu}\,.
\end{equation}
%

\section{Details of computation of the UV amplitudes}
\label{app:UV}
%
%
\subsection{Amplitude for one $B$ and no Higgs bosons}
%
This amplitude in the UV is given by the diagrams $(a)$ and $(b)$ in Fig.~\ref{fig:NNB_and_vNH}. We have:
\begin{align}
 i\mathcal{M}_{UV}^a = -g^\prime g_N^2 \overline{u}(p_N-p_B)P_L
 \bigg\lbrace& \mu^{4-d} \int \frac{\mathrm{d}^{d}k}{(2\pi)^{d}} \frac{1}{D^3} (\slashed{p}_N-\slashed{p}_B+\slashed{k}+M) \gamma^\mu (\slashed{p}_N+\slashed{k}+M) \nonumber \\
 &\times\left[1-\frac{2(p_N-p_B)k+(p_N-p_B)^2}{D}+\frac{4[k(p_N-p_B)]^2}{D^2}\right] \nonumber \\
 &\times\left[1-\frac{2p_N k+p_N^2}{D}+\frac{4(kp_N)^2}{D^2}\right]\bigg\rbrace P_R u(p_N)\epsilon_\mu^*(p_B) \nonumber \\
 = -g^\prime g_N^2  \overline{u}(p_N-p_B)P_L\bigg\lbrace& 
 [\mu^{2\epsilon}(2\epsilon -2)B_3 + M^2 A_3]\gamma^\mu 
 \nonumber \\
 &+ \left[12 B_4-A_3-48 C_5-2 M^2 A_4+12 M^2 B_5\right] \gamma^\mu p_N^2 \nonumber \\
 &+ \left[4 B_4 - 16 C_5 - M^2 A_4 + 4 M^2 B_5\right] \gamma^\mu p_B^2 \nonumber \\
 &+ \left[48 C_5 - 8 B_4 + 2 M^2 A_4 - 12 M^2 B_5\right] \gamma^\mu p_B p_N
 + \left[A_3 - 4 B_4\right] \gamma^\mu \slashed{p}_B \slashed{p}_N \nonumber \\
 &+ \left[2 A_3 - 16 B_4 + 48 C_5\right] p_N^\mu \slashed{p}_N 
 + \left[16 C_5 - 4 B_4\right] p_B^\mu \slashed{p}_B \nonumber \\
 &+ \left[12 B_4 - 2 A_3 - 24 C_5\right] p_B^\mu \slashed{p}_N 
 + \left[4 B_4 - 24 C_5\right] p_N^\mu \slashed{p}_B
 \bigg\rbrace 
 u(p_N)\epsilon_\mu^*(p_B)\nonumber\\
 = \frac{ig^\prime g_N^2}{192\pi^2 M^2} \overline{u}(p_N-p_B)P_L 
 \bigg\lbrace& \gamma^\mu \left(6 M^2\log\frac{\mu^2}{M^2}
 +  p_N^2 + 3 p_B^2 - 3 p_B p_N + 2 \slashed{p}_B \slashed{p}_N \right) \nonumber \\
 &+ 2 p_N^\mu \slashed{p}_N - 2 p_B^\mu \slashed{p}_B 
 - 3 p_B^\mu \slashed{p}_N + p_N^\mu \slashed{p}_B
 \bigg\rbrace u(p_N) \epsilon_\mu^*(p_B)\,.
\end{align}
Here and in what follows $D \equiv k^2 - M^2$. 
The second diagram leads to
\begin{align}
 i\mathcal{M}_{UV}^b = -g^\prime g_N^2 \overline{u}(p_N-p_B) P_L
 \bigg\lbrace& \mu^{4-d} \int \frac{\mathrm{d}^{d}k}{(2\pi)^d} 
 \frac{1}{D^3} 
 (\slashed{k}+\slashed{p}_N+M) (2 k^\mu+p_B^\mu) \nonumber \\
 &\times\left[1-\frac{2p_Nk+p_N^2}{D}+\frac{4(kp_N)^2}{D^2}\right] \nonumber \\
 &\times\left[1-\frac{2p_B k+p_B^2}{D}+\frac{4(kp_B)^2}{D^2}\right]\bigg\rbrace P_R u(p_N)\epsilon_\mu^*(p_B)\nonumber \\
 = -g^\prime g_N^2  \overline{u}(p_N - p_B)P_L
 \bigg\lbrace& 2\mu^{2\epsilon}B_3\gamma^\mu 
 + \left[8 C_5 - 2 B_4\right] \gamma^\mu p_N^2 
 + \left[8 C_5 - 2 B_4\right] \gamma^\mu p_B^2 \nonumber \\
 &+ 8 C_5 \gamma^\mu p_B p_N 
 +\left[16 C_5 - 4 B_4\right] p_N^\mu \slashed{p}_N 
 +\left[16 C_5 - 2 B_4\right] p_B^\mu \slashed{p}_B \nonumber \\
 &+\left[A_3 - 6 B_4 + 8 C_5\right] p_B^\mu \slashed{p}_N 
 + 8 C_5 p_N^\mu \slashed{p}_B
 \bigg\rbrace u(p_N)\epsilon_\mu^*(p_B)\nonumber\\
 = \frac{ig^\prime g_N^2}{192\pi^2 M^2} \overline{u}(p_N - p_B)P_L
 \bigg\lbrace& \gamma^\mu 
 \left(-6 M^2 \log\frac{\mu^2}{M^2} - p_N^2 - p_B^2 + p_B p_N\right) \nonumber\\
 &- 2 p_N^\mu \slashed{p}_N + p_B^\mu \slashed{p}_N + p_N^\mu \slashed{p}_B 
 \bigg\rbrace u(p_N)\epsilon^*_\mu(p_B)\,.
\end{align}
Adding the two pieces together:
\begin{align}
 i\mathcal{M}_{UV} = i\mathcal{M}_{UV}^{a} + i\mathcal{M}_{UV}^{b}
 \phantom{{}===={}}& \nonumber \\
 = \frac{ig^\prime g_N^2}{96\pi^2 M^2} 
 \overline{u}(p_N-p_B)P_L
 \bigg\lbrace& \gamma^\mu 
 \left(p_B^2 - p_B p_N + \slashed{p}_B \slashed{p}_N\right) \nonumber\\
 &- p_B^\mu \slashed{p}_B - p_B^\mu \slashed{p}_N + p_N^\mu \slashed{p}_B 
 \bigg\rbrace u(p_N)\epsilon^*_\mu(p_B)\,.
\end{align}

\subsection{Amplitude for one Higgs and no gauge bosons}
%
This UV amplitude is represented by the diagram $(c)$ 
in Fig.~\ref{fig:NNB_and_vNH}. We have:
\begin{align}
 i\mathcal{M}_{UV} = -\frac{g_N g_X g_L}{\sqrt{2}}\overline{u}(p_N) P_L 
 \Bigg\lbrace& \mu^{4-d} \int \frac{\mathrm{d}^{d}k}{(2\pi)^{d}} \frac{1}{D^3} 
 \left(\slashed{p}_{N}+\slashed{k}+M\right)
 \left(\slashed{p}_\nu + \slashed{k}+M\right) \nonumber \\
 &\times \left[1-\frac{2kp_N+p_N^2}{D}+\frac{4(kp_N)^2}{D^2}\right] \nonumber \\
 &\times \left[1-\frac{2kp_\nu+p_\nu^2}{D}+\frac{4(kp_\nu)^2}{D^2}\right] 
 \Bigg\rbrace P_L u(p_\nu) \nonumber \\
 = -\frac{g_N g_X g_L}{\sqrt{2}}\overline{u}(p_N) P_L 
 \bigg\lbrace& \mu^{2\epsilon} (4-2\epsilon) B_3 + M^2 A_3 \nonumber \\
 &-\left[6\left(B_4-4C_5\right) + M^2\left(A_4-4B_5\right)\right] \left(p_\nu^2+p_N^2\right) \nonumber \\
 &+ 4\left(6C_5+M^2B_5\right)p_{\nu}p_N 
 + \left(A_3-4B_4\right)\slashed{p}_N\slashed{p}_\nu \bigg\rbrace u(p_\nu) \nonumber \\
 = \frac{i g_N g_X g_L}{96\sqrt{2}\pi^2M^2}\overline{u}(p_N) P_L 
 &\bigg\lbrace 6M^2\left(1-\log{\frac{\mu^2}{M^2}}\right) - p_\nu^2 - p_N^2 + p_{\nu}p_N 
 + \slashed{p}_N\slashed{p}_\nu \bigg\rbrace u(p_\nu)\,.
\end{align}

\subsection{Amplitude for one Higgs and one photon}
%
The relevant UV diagrams are depicted in Fig.~\ref{fig:vNAH}. 
We have:
\begin{align}
 i\mathcal{M}_{UV}^{a+b} = \frac{g_L g_X g_N e}{\sqrt{2}} \overline{u}(p_N)P_L
 & \int \frac{\mathrm{d}^4k}{(2\pi)^4} \frac{1}{D^4}
 \left(\slashed{p}_{N}+\slashed{k}+M\right) \left[1-\frac{2kp_N}{D}\right]  \nonumber \\
 &\times\Bigg\lbrace \gamma^\mu
 \left(\slashed{p}_{\gamma}+\slashed{p}_{N}+\slashed{k}+M\right) 
 \left[1-\frac{2k(p_\gamma+p_N)}{D}\right] \nonumber \\
 &\phantom{{}\times{}}
 +\left(\slashed{p}_{h}+\slashed{p}_{N}+\slashed{k}+M\right) \gamma^\mu
 \left[1-\frac{2k(p_h+p_N)}{D}\right] \Bigg\rbrace \nonumber \\
 &\times\left(\slashed{p}_{h}+\slashed{p}_{\gamma}+\slashed{p}_{N}+\slashed{k}+M\right) 
 \left[1-\frac{2k(p_h+p_\gamma+p_N)}{D}\right] \nonumber \\
 &\times P_L u(p_\nu) \epsilon_\mu^*(p_\gamma) \nonumber \\
 =\frac{g_L g_X g_N e}{\sqrt{2}} \overline{u}(p_N)P_L \bigg\lbrace 
 &2 \left[2 B_4 - 12 C_5 + M^2 \left(A_4 - 10 B_5\right)\right]p_h^\mu \nonumber \\
 &+ 4 \left[B_4 - 12 C_5 - 4 M^2 B_5\right]p_\gamma^\mu \nonumber \\
 &+6 \left[2 B_4 - 12 C_5 + M^2 \left(A_4 - 6 B_5\right) \right] p_N^\mu \nonumber \\
 &+ \left[2 B_4 - 12 C_5 + M^2 \left(A_4 +2 B_5\right)\right]\gamma^\mu \slashed{p}_h \nonumber \\
 &+ \left[2 B_4 + 12 C_5 + M^2 \left(3 A_4 - 2 B_5\right)\right]\gamma^\mu \slashed{p}_\gamma \bigg\rbrace u(p_\nu) \epsilon_\mu^*(p_\gamma) \nonumber \\
 = \frac{i g_L g_X g_N e}{96\sqrt{2}\pi^2M^2} \overline{u}(p_N)P_L \bigg\lbrace
 &- p_h^\mu - p_\gamma^\mu + \gamma^\mu \slashed{p}_h + \gamma^\mu \slashed{p}_\gamma \bigg\rbrace 
 u(p_\nu)\epsilon_\mu^*(p_\gamma)\,,
 \label{eq:1H1gammaUVab}
\end{align}
and
\begin{align}
 i\mathcal{M}_{UV}^{c} = \frac{g_L g_X g_N e}{\sqrt{2}} \overline{u}(p_N)P_L 
 \Bigg\lbrace& \int \frac{\mathrm{d}^4k}{(2\pi)^4} \frac{1}{D^4} 
 \left(\slashed{p}_{\gamma}+\slashed{p}_{N}+\slashed{k}+M\right)
 \left(\slashed{p}_{h}+\slashed{p}_{\gamma}+\slashed{p}_{N}+\slashed{k}+M\right) \nonumber \\
 &\times\left(p_\gamma^\mu + 2 k^\mu\right) 
 \left[1-\frac{2k(p_\gamma+p_N)}{D}\right] 
 \left[1-\frac{2k(p_h+p_\gamma+p_N)}{D}\right] \nonumber \\
 &\times\left[1-\frac{2kp_\gamma}{D}\right] \Bigg\rbrace 
 P_L u(p_\nu) \epsilon_\mu^*(p_\gamma)\nonumber \\
 = \frac{g_L g_X g_N e}{\sqrt{2}} \overline{u}(p_N)P_L 
 \bigg\lbrace& 
 -2 \left[12 C_5+2 M^2 B_5\right] p_h^\mu + 2 B_4 \gamma^\mu \slashed{p}_h \nonumber \\
 &+ \left[8 \left(B_4 - 9 C_5\right)+M^2\left(A_4 - 12 B_5\right)\right] p_\gamma^\mu \nonumber \\
 &+ 4\left[B_4-12 C_5-2 M^2 B_5\right] p_N^\mu 
 \bigg\rbrace u(p_\nu) \epsilon_\mu^*(p_\gamma) \nonumber \\
 = \frac{i g_L g_X g_N e}{96\sqrt{2}\pi^2M^2} \overline{u}(p_N)P_L\bigg\lbrace&
 p_h^\mu - \gamma^\mu \slashed{p}_h \bigg\rbrace u(p_\nu) \epsilon_\mu^*(p_\gamma)\,.
 \label{eq:1H1gammaUVc}
\end{align}
Adding Eqs.~\eqref{eq:1H1gammaUVab} and 
\eqref{eq:1H1gammaUVc} together, we get:
\begin{equation}
 i\mathcal{M}_{UV} = i\mathcal{M}_{UV}^{a+b}+i\mathcal{M}_{UV}^c 
 =  \frac{i g_L g_X g_N e}{96\sqrt{2}\pi^2 M^2}\overline{u}(p_N)P_L 
 \bigg\lbrace\gamma^\mu\slashed{p}_\gamma - p_\gamma^\mu\bigg\rbrace u(p_\nu) \epsilon_\mu^*(p_\gamma)\,.
\end{equation}
%

\subsection{Amplitude for one Higgs and one $W$}
%
This amplitude in the UV is depicted by the diagram $(a)$ 
in Fig.~\ref{fig:eNWH_and_NNHH}. We have:
\begin{align}
 i\mathcal{M}_{UV} = \frac{g_N g_X g_L g}{2}
 \overline{u}(p_N) P_L \Bigg\lbrace& \int \frac{\mathrm{d}^4 k}{(2\pi)^4} \frac{1}{D^4}
 \left(\slashed{p}_{N}+\slashed{k}+M\right)
 \left(\slashed{p}_h+\slashed{p}_N+\slashed{k}+M\right) \gamma^\mu \nonumber \\
 &\times\left(\slashed{p}_W+\slashed{p}_h+\slashed{p}_N+\slashed{k}+M\right) 
 \left[1-\frac{2kp_N}{D}\right] \nonumber \\
 &\times\left[1-\frac{2k(p_h+p_N)}{D}\right] 
 \left[1-\frac{2k(p_W+p_h+p_N)}{D}\right] \Bigg\rbrace\nonumber \\
 &\times P_L u(p_e) \epsilon_\mu^\ast(p_W) \nonumber \\
 = \frac{g_N g_X g_L g}{2} 
 \overline{u}(p_N) P_L \bigg\lbrace& 4M^2\left(A_4 - 6B_5\right) p_N^\mu 
 +\left[4B_4 + 2M^2\left(A_4-8B_5\right)\right] p_h^\mu \nonumber \\
 &-8M^2B_5 p_W^\mu 
 +\left[6\left(B_4-6C_5\right) - M^2\left(A_4-6B_5\right)\right] \gamma^\mu \slashed{p}_N \nonumber \\
 &+4\left(B_4-6C_5+M^2B_5\right) \gamma^\mu \slashed{p}_h \nonumber \\
 &+\left[4\left(B_4-3C_5\right) + M^2\left(A_4+2B_5\right)\right] \gamma^\mu \slashed{p}_W
 \bigg\rbrace u(p_e) \epsilon_\mu^\ast(p_W) \nonumber \\
 = \frac{i g_N g_X g_L g}{192 \pi^2 M^2} 
 \overline{u}(p_N) P_L \bigg\lbrace& p_N^\mu - 2p_h^\mu - p_W^\mu - \gamma^\mu \slashed{p}_N \bigg\rbrace u(p_e) \epsilon_\mu^\ast(p_W)\,.
\end{align}
%

\subsection{Amplitude for two Higgses and no gauge bosons}
%
This UV amplitude is given by the diagrams $(b)$ and $(c)$ 
in Fig.~\ref{fig:eNWH_and_NNHH}. Taking into account 
possible permutations of $p_h$ and $p_{h'}$, we have:
\begin{align}
 i\mathcal{M}_{UV}^{b}
 = \frac{g_N^2 g_X^2}{2} \overline{u}(p_{N'}) P_L 
 & \int \frac{\mathrm{d}^4 k}{(2\pi)^4} \frac{1}{D^4}
 \left(\slashed{p}_{N'}+\slashed{k}+M\right) \left[1-\frac{2kp_{N'}}{D}\right] \nonumber \\
 &\times\Bigg\lbrace
 \left(\slashed{p}_{N}-\slashed{p}_{h}+\slashed{k}+M\right) 
 \left[1-\frac{2kp_N}{D}+\frac{2kp_h}{D}\right] \nonumber \\
 &\phantom{{}\times{}}+\left(\slashed{p}_{N}-\slashed{p}_{h'}+\slashed{k}+M\right) 
 \left[1-\frac{2kp_N}{D}+\frac{2k p_{h'}}{D}\right] \Bigg\rbrace\nonumber \\
 &\times \left(\slashed{p}_{N}+\slashed{k}+M\right) 
 \left[1-\frac{2kp_N}{D}\right] P_R u(p_N) \nonumber \\
 = \frac{g_N^2 g_X^2}{2} \overline{u}(p_{N'}) P_L \bigg\lbrace&
 2\left[2 B_4 - 24 C_5 + 2 M^2 A_4 - 12 M^2 B_5\right] \slashed{p}_N \nonumber \\
 &+\left[2 B_4 + 12 C_5 - M^2 A_4 + 6 M^2 B_5\right] \slashed{p}_h \nonumber \\
 &+\left[2 B_4 + 12 C_5 - M^2 A_4 + 6 M^2 B_5\right] \slashed{p}_{h'} \nonumber \\
 &+2\left[4 B_4 - 12 C_5 + M^2 A_4 - 6 M^2 B_5\right] \slashed{p}_{N'}
 \bigg\rbrace u(p_N) \nonumber \\
 = \frac{ig_N^2 g_X^2}{96\pi^2M^2} \overline{u}(p_{N'}) P_L \bigg\lbrace&
 \slashed{p}_N - \slashed{p}_h - \slashed{p}_{h'} - \slashed{p}_{N'} 
 \bigg\rbrace u(p_N) = 0\,,
\end{align}
by virtue of the momentum conservation. Thus, we get:
\begin{align}
 i\mathcal{M}_{UV} = i\mathcal{M}_{UV}^{c} = g_N^2 \lambda_{\varphi H} \overline{u}(p_{N'}) P_L 
 \Bigg\lbrace& \int \frac{\mathrm{d}^4 k}{(2\pi)^4} \frac{1}{D^3}
 \left(\slashed{p}_{N}+\slashed{k}+M\right) \nonumber \\
 &\times \left[1-\frac{2kp_N}{D}\right]
 \left[1-\frac{2k(p_N-p_{N'})}{D}\right] 
 \Bigg\rbrace P_R u(p_N) \nonumber \\
 = g_N^2 \lambda_{\varphi H} \overline{u}(p_{N'}) P_L \bigg\lbrace&
 \left[A_3 - 4 B_4\right] \slashed{p}_N 
 + 2 B_4 \slashed{p}_{N'} \bigg\rbrace u(p_N) \nonumber \\
 = - \frac{ig_N^2 \lambda_{\varphi H}}{96\pi^2M^2} 
 \overline{u}(p_{N'}) P_L \bigg\lbrace&\slashed{p}_N + \slashed{p}_{N'}\bigg\rbrace u(p_N)\,.
\end{align}
%

\subsection{Amplitude for two Higgses and one $W^3$}
%
The relevant diagram in the UV is  the diagram $(a)$ 
in Fig.~\ref{fig:NNHHW3_and_vNHHH}. We have:
\begin{align}
 i\mathcal{M}_{UV} &= -\frac{g g_N^2 g_{X}^2}{2} \overline{u}(p_{N'}) P_L\int \frac{\mathrm{d}^4 k}{(2\pi)^4}\frac{1}{D^5} 
 \left(\slashed{k}+M\right)^2 \gamma^\mu \left(\slashed{k}+M\right)^2 P_R u(p_N) \epsilon_\mu^*(p_{W^3}) \nonumber \\
 &= -\frac{g g_N^2 g_{X}^2}{2} \overline{u}(p_{N'}) P_L \left(24 C_5 + M^4 A_5\right) \gamma^\mu u(p_N) \epsilon_\mu^*(p_{W^3}) \nonumber \\
 &= \frac{i g g_N^2 g_{X}^2}{96\pi^2 M^2} \overline{u}(p_{N'}) P_L \gamma^\mu u(p_N) \epsilon_\mu^*(p_{W^3})\,.
\end{align}
%

\subsection{Amplitude for three Higgses and no gauge bosons}
%
This UV amplitude is represented by the diagrams $(b)$ and $(c)$ 
in Fig.~\ref{fig:NNHHW3_and_vNHHH}. We have:
\begin{align}
 i\mathcal{M}_{UV}^{b}
 &= -\frac{3 g_N g_X^3 g_L}{\sqrt{2}}\overline{u}(p_{N})P_L 
 \int \frac{\mathrm{d}^4 k}{(2\pi)^4}\frac{1}{D^5} \left(\slashed{k} + M\right)^4 
 P_L u(p_\nu) \nonumber \\
 &= -\frac{3 g_N g_X^3 g_L}{\sqrt{2}} \overline{u}(p_N) P_L 
 \left(24 C_5 + 24 M^2 B_5 + M^4 A_5\right) u(p_\nu) \nonumber \\
 &= - \frac{i g_N g_X^3 g_L}{32\sqrt{2}\pi^2M^2} \overline{u}(p_N) P_L  u(p_\nu)\,,
\end{align}
\begin{align}
 i\mathcal{M}_{UV}^{c}
 &= -\frac{3 \lambda_{\varphi H} g_N g_X g_L}{\sqrt{2}} 
 \overline{u}(p_{N})P_L 
 \int \frac{\mathrm{d}^4 k}{(2\pi)^4}\frac{1}{D^4} \left(\slashed{k} + M\right)^2 
 P_L u(p_\nu) \nonumber \\
 &= -\frac{3 \lambda_{\varphi H} g_N g_X g_L}{\sqrt{2}} \overline{u}(p_N) P_L 
 \left(4 B_4 + M^2 A_4\right) u(p_\nu) \nonumber \\
 &=\frac{i \lambda_{\varphi H} g_N g_X g_L}{32\sqrt{2}\pi^2M^2} 
 \overline{u}(p_N) P_L  u(p_\nu)\,.
\end{align}
Finally,
\begin{equation}
 i\mathcal{M}_{UV} = i\mathcal{M}_{UV}^{b} + i\mathcal{M}_{UV}^{c} 
 =\frac{i g_N g_X g_L}{32\sqrt{2}\pi^2M^2} 
 \left(\lambda_{\varphi H} - g_X^2\right) 
 \overline{u}(p_N) P_L  u(p_\nu)\,.
\end{equation}
%

\subsection{Amplitude for four $N$ fermions}
%
This UV amplitude is depicted by the diagram $(a)$ 
in Fig.~\ref{fig:NNNN_and_NNvv} 
(note that there is a second diagram with opposite sign 
due to the exchange of identical fermions). We have:
\begin{align}
 i\mathcal{M}_{UV} = g_N^4\int\frac{\mathrm{d}^4 k}{(2\pi)^4}\frac{1}{D^4}\bigg\lbrace&
 \left[\overline{u}(p_3)P_L\left(\slashed{k}+M\right)P_R u(p_1)\right]
 \left[\overline{u}(p_4)P_L\left(\slashed{k}+M\right)P_R u(p_2)\right] \nonumber \\
 &- \left[\overline{u}(p_4)\left(\slashed{k}+M\right)u(p_1)\right]
 \left[\overline{u}(p_3)\left(\slashed{k}+M\right)P_R u(p_2)\right]
 \bigg\rbrace \nonumber \\
 = g_N^4 B_4 \bigg\lbrace& 
 \left[\overline{u}(p_3) \gamma^\mu P_R u(p_1)\right]
 \left[\overline{u}(p_4)\gamma_\mu P_R u(p_2)\right] \nonumber \\
 & -[\overline{u}(p_4)\gamma^\mu P_R u(p_1)][\overline{u}(p_3)\gamma_\mu P_R u(p_2)]\bigg\rbrace \nonumber \\
 = 2 g_N^4 B_4& 
 \left[\overline{u}(p_3)\gamma^\mu P_R u(p_1)\right]
 \left[\overline{u}(p_4)\gamma_\mu P_R u(p_2)\right] \nonumber \\
 = - \frac{i g_N^4}{96 \pi^2 M^2}&
 \left[\overline{u}(p_3)\gamma^\mu P_R u(p_1)\right]
 \left[\overline{u}(p_4)\gamma_\mu P_R u(p_2)\right].
\end{align}
In the
penultimate step, we have rearranged the spinors using 
a Fierz identity.

\subsection{Amplitude for two $N$ fermions and two neutrinos}
%
This amplitude in the UV is given by the diagram $(b)$ 
in Fig.~\ref{fig:NNNN_and_NNvv}. We have:
\begin{align}
 i\mathcal{M}_{UV} &= g_N^2 g_L^2\int\frac{\mathrm{d}^4 k}{(2\pi)^4}\frac{1}{D^4}
 \left[\overline{u}(p_{\nu'})P_R\left(\slashed{k}+M\right)P_L u(p_\nu)\right]
 \left[\overline{u}(p_{N'})P_L\left(\slashed{k}+M\right)P_R u(p_N)\right] \nonumber\\
 &=g_N^2 g_L^2B_4 
 \left[\overline{u}(p_{\nu'})\gamma^\mu P_L u(p_\nu)\right]
 \left[\overline{u}(p_{N'})\gamma_\mu P_R u(p_N)\right] \nonumber \\
 &= -\frac{ig_N^2 g_L^2}{192\pi^2 M^2} 
 \left[\overline{u}(p_{\nu'})\gamma^\mu P_L u(p_\nu)\right]
 \left[\overline{u}(p_{N'})\gamma_\mu P_R u(p_N)\right].
\end{align}
%

\section{Matching of renormalisable terms}
\label{app:dim4}
%
%
\subsection*{Corrections to the Higgs propagator}
\begin{figure}[t]
 \centering
 \includegraphics[width=\textwidth]{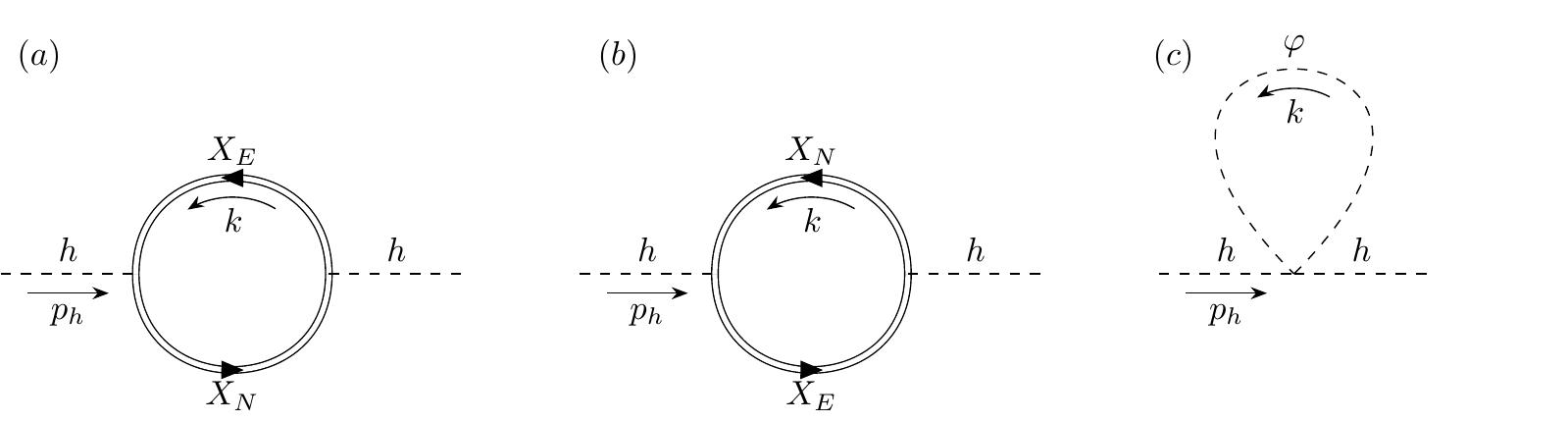}
 \caption{\it Diagrams contributing to the Higgs self-energy in the UV.}
 \label{fig:H_self-energy}
\end{figure}

%
The diagrams $(a)$ and $(b)$ in Fig.~\ref{fig:H_self-energy} 
lead to the following contribution to the Higgs self-energy:
\begin{align}
 -i\left(M^{2}_{UV}\right)^{a+b} &= - g_X^2
 \mu^{4-d} \int \frac{\mathrm{d}^{d}k}{(2\pi)^{d}} 
 \frac{1}{D^2} \Bigg\lbrace
 \tr\left[\left(\slashed{k}+\slashed{p}_h+M\right)
 \left(\slashed{k}+M\right)\right] \nonumber \\
 &\hspace{4.6cm}\times\left[1-\frac{2kp_h+p_h^2}{D}+\frac{4(kp_h)^2}{D^2}\right] 
 \Bigg\rbrace  \nonumber \\
 &= -4g_X^2 \mu^{2\epsilon} \bigg\lbrace
 (4-2\epsilon)B_2 + M^2 A_2 \nonumber \\
 &\hspace{2.3cm}+ \left[(6-2\epsilon)\left(4C_4 - B_3\right) + M^2 \left(4B_4 - A_3\right)\right] p_h^2
 \bigg\rbrace\nonumber\\
 &=-\frac{i g_X^2}{12 \pi^2} \bigg\lbrace
 3\left(1 + 3\log\frac{\mu^2}{M^2}\right) M^2
 + \left(1 - 3\log\frac{\mu^2}{M^2}\right) p_h^2 \bigg\rbrace\,.
 \end{align}
The diagram $(c)$ in Fig.~\ref{fig:H_self-energy} gives
\begin{align}
 -i \left(M^2_{UV}\right)^c = \lambda_{\varphi H} \mu^{4-d} 
 \int \frac{\mathrm{d}^{d}k}{(2\pi)^{d}} \frac{1}{D} 
 = \lambda_{\varphi H} \mu^{2\epsilon} A_1 
 = \frac{i \lambda_{\varphi H}}{16 \pi^2} \left(1+\log\frac{\mu^2}{M^2}\right)M^2\,.
 \end{align}
Finally, their sum yields
\begin{align}
-i M^2_{UV} = -i \left(M^2_{UV}\right)^{a+b} -i \left(M^2_{UV}\right)^c 
&= -\frac{i}{48\pi^2} \left[ 3\left(4g_X^2 - \lambda_{\varphi H}\right) M^2 +4g_X^2p_h^2 \right] \nonumber \\
&\phantom{{}={}}-\frac{i}{16\pi^2} \left[\left(12g_X^2-\lambda_{\varphi H}\right)M^2 - 4g_X^2p_h^2\right] \log\frac{\mu^2}{M^2}\,.
\label{eq:MsqUV}
\end{align}
%

\subsection*{Corrections to the fermion propagators}
\begin{figure}[t]
 \centering
 \includegraphics[width=0.75\textwidth]{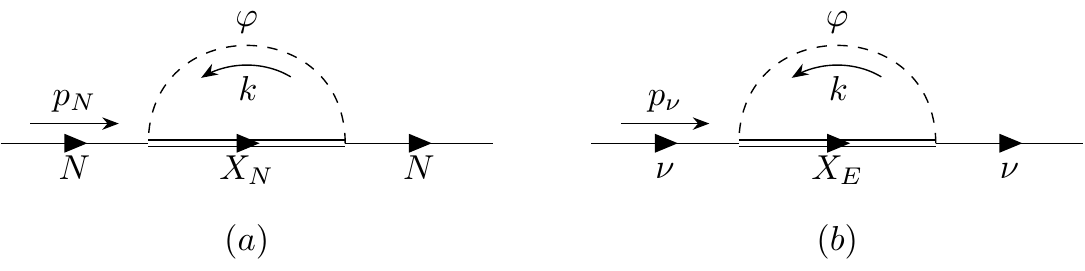}
 \caption{\it Diagrams contributing to the $N$ and $\nu$ self-energies in the UV.}
 \label{fig:N_and_v_self-energies}
\end{figure}

%
For the contribution to the self-energy of $N$ depicted by the diagram $(a)$ 
in Fig.~\ref{fig:N_and_v_self-energies},
we find
\begin{align}
 -i\Sigma_{UV} &= g_N^2 P_L \mu^{4-d} 
 \int \frac{\mathrm{d}^{d}k}{(2\pi)^{d}} \frac{1}{D^2}
 (\slashed{k}+\slashed{p}_N+M) 
 \left[1-\frac{2kp_N}{D}\right] P_R \nonumber \\
 &= g_N^2 \mu^{2\epsilon} \left(A_2-2B_3\right) P_L \slashed{p}_N
 =\frac{i g_N^2}{32\pi^2} \log\frac{\mu^2}{M^2} P_L\slashed{p}_N\,.
\end{align} 
A similar contribution to the neutrino self-energy 
represented by the diagram $(b)$ in Fig.~\ref{fig:N_and_v_self-energies} reads
\begin{align}
 -i\Sigma_{UV} &= g_L^2 P_R \mu^{4-d} 
 \int \frac{\mathrm{d}^{d}k}{(2\pi)^{d}} \frac{1}{D^2}
 (\slashed{k}+\slashed{p}_\nu+M) 
 \left[1-\frac{2kp_\nu}{D}\right] P_L \nonumber \\
 &= g_L^2 \mu^{2\epsilon} \left(A_2-2B_3\right) P_R \slashed{p}_\nu
 =\frac{i g_L^2}{32\pi^2} \log\frac{\mu^2}{M^2} P_R\slashed{p}_\nu\,.
\end{align} 
%

\subsection*{Corrections to the gauge boson propagators}
\begin{figure}[t]
 \centering
 \includegraphics[width=\textwidth]{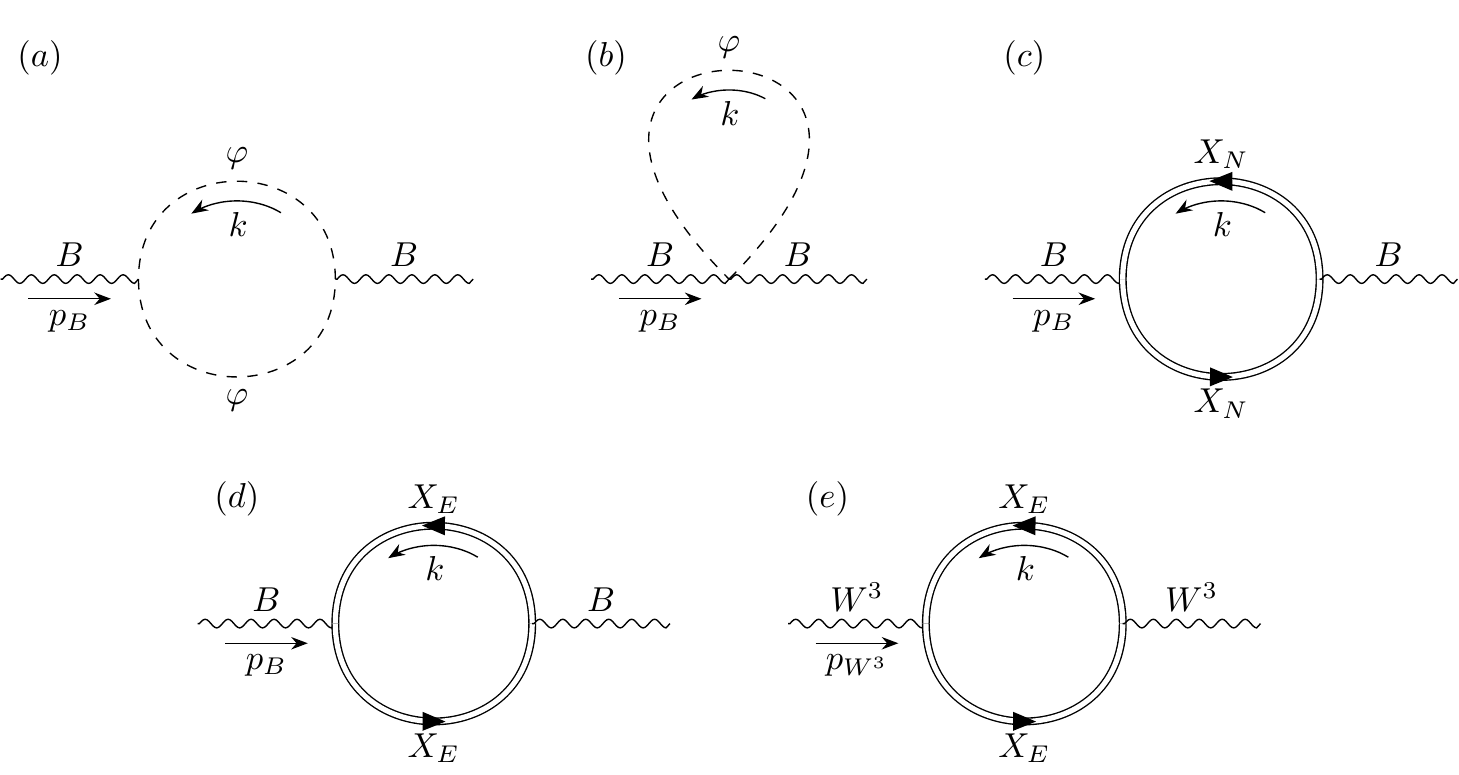}
 \caption{\it Diagrams contributing to the $B$ and $W^3$ self-energies in the UV.}
 \label{fig:B_and_W3_self-energies}
\end{figure}

%
There are four diagrams contributing to the self-energy of $B_\mu$, see Fig.~\ref{fig:B_and_W3_self-energies}. 
The diagram~$(a)$ gives
\begin{align}
 i\left(\Pi^{\mu\nu}_{UV}\right)^a &= g^{\prime2} \mu^{4-d} 
 \int \frac{\mathrm{d}^{d}k}{(2\pi)^{d}} \frac{1}{D^2} 
 \left(2k^\mu + p_B^\mu\right) \left(2k^\nu + p_B^\nu\right) 
 \left[1 - \frac{2kp_B+p_B^2}{D} + \frac{4(kp_B)^2}{D^2}\right] \nonumber \\
 &=g^{\prime2} \mu^{2\epsilon} \left[
 4B_2 g^{\mu\nu} + \left(16C_4 - 4B_3\right) p_B^2 g^{\mu\nu} 
 + \left(A_2 - 8B_3 + 32C_4\right) p_B^\mu p_B^\nu\right] \nonumber \\
 &=\frac{ig^{\prime2}}{48\pi^2}
 \left[6M^2\left(1+\log\frac{\mu^2}{M^2}\right)g^{\mu\nu} 
 - \log\frac{\mu^2}{M^2}\left(p_B^2 g^{\mu\nu} - p_B^\mu p_B^\nu\right)\right].
\end{align}
The diagram~$(b)$ yields
\begin{align}
 i\left(\Pi^{\mu\nu}_{UV}\right)^b &= -2g^{\prime2} g^{\mu\nu} \mu^{4-d} 
 \int \frac{\mathrm{d}^{d}k}{(2\pi)^{d}} \frac{1}{D} \nonumber \\
 &= -2g^{\prime2} g^{\mu\nu} \mu^{2\epsilon} A_1 
 = -\frac{ig^{\prime2}}{8\pi^2} M^2\left(1+\log\frac{\mu^2}{M^2}\right)g^{\mu\nu}\,.
\end{align}
Further, for the diagram ($c$) we find
\begin{align}
 i\left(\Pi^{\mu\nu}_{UV}\right)^c &= -g^{\prime2} 
 \mu^{4-d} \int \frac{\mathrm{d}^{d}k}{(2\pi)^{d}} \Bigg\lbrace\frac{1}{D^2} 
 \tr\left[\gamma^\nu\left(\slashed{k}+\slashed{p}_B+M\right)
 \gamma^\mu\left(\slashed{k}+M\right)\right] \nonumber \\
 &\hspace{4cm}\times\left[1 - \frac{2kp_B+p_B^2}{D} + \frac{4(kp_B)^2}{D^2}\right] \Bigg\rbrace \nonumber \\
 &= -4g^{\prime2} \mu^{2\epsilon} \bigg\lbrace 
 \left[M^2 A_2 + (2\epsilon-2)B_2\right] g^{\mu\nu} 
 +\left[16C_4 - 4B_3\right] p_B^\mu p_B^\nu \nonumber \\
 &\hspace{2.25cm}+\left[(4-2\epsilon)\left(B_3-4C_4\right) - M^2\left(A_3-4B_4\right)\right] p_B^2 g^{\mu\nu} \bigg\rbrace \nonumber \\
 &=-\frac{ig^{\prime2}}{12\pi^2} \log\frac{\mu^2}{M^2} 
 \left(p_B^2 g^{\mu\nu} - p_B^\mu p_B^\nu\right).
\label{eq:PiBXN}
\end{align}
Finally, the diagram~$(d)$ leads 
to the same result divided by 4 because of $Y_{X_E} = 1/2$ and multiplied 
by 2 because both $X_E^+$ and $X_E^0$ contribute. 
Summing all contributions we obtain
\begin{equation}
 i\Pi^{\mu\nu}_{UV} = - \frac{7ig^{\prime2}}{48\pi^2} \log\frac{\mu^2}{M^2} \left(p_B^2 g^{\mu\nu} - p_B^\mu p_B^\nu\right).
\end{equation}

 Computation of the diagram~$(e)$ in Fig.~\ref{fig:B_and_W3_self-energies} 
providing a contribution to the $W^3_\mu$ self-energy is almost the same 
as that in Eq.~\eqref{eq:PiBXN}. 
As a result, we find
\begin{equation}
i\Pi^{\mu\nu}_{UV} = - \frac{ig^2}{24\pi^2} \log\frac{\mu^2}{M^2} \left(p_{W^3}^2 g^{\mu\nu} - p_{W^3}^\mu p_{W^3}^\nu\right).
\end{equation}
%

\subsection*{Corrections to the Higgs quartic coupling}
\begin{figure}[t]
 \centering
 \includegraphics[width=0.66\textwidth]{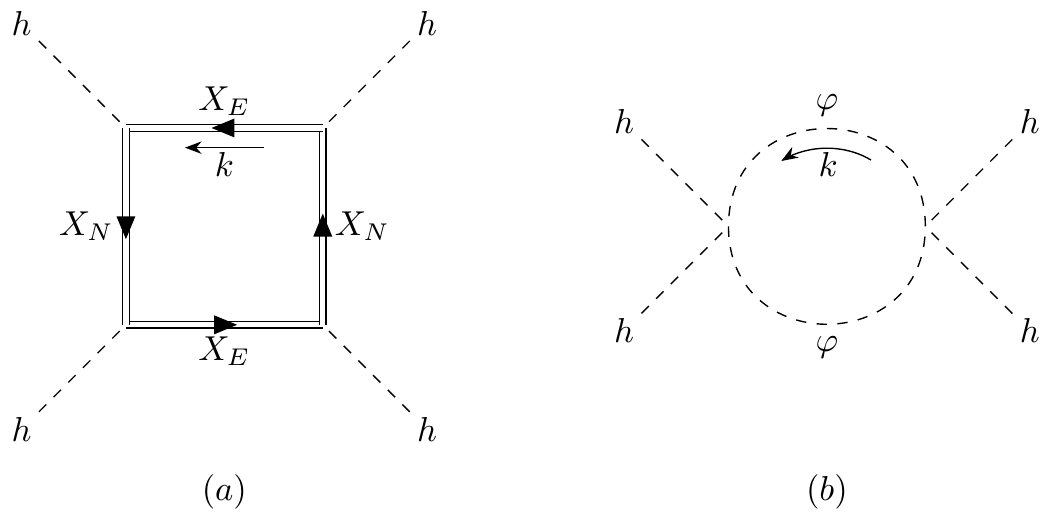}
 \caption{\it Diagrams contributing to the amplitude $\langle hhhh\rangle$ in the UV.}
 \label{fig:HHHH}
\end{figure}

%
The diagram~$(a)$ in Fig.~\ref{fig:HHHH} 
reads
\begin{align}
 i\mathcal{M}_{UV}^a &= - 12 \frac{g_X^4}{4} \mu^{4-d} 
 \int \frac{\mathrm{d}^{d}k}{(2\pi)^{d}} 
 \frac{\tr[(\slashed{k}+M)^4]}{D^4} \nonumber \\
 &= - 3 g_X^4 \left[M^4A_4 + 24M^2B_4 
 + \mu^{2\epsilon}(4-2\epsilon)(6-2\epsilon)C_4\right] \nonumber \\
 &= \frac{2 i g_X^4}{\pi^2} \left(1 - \frac{3}{8} 
\log\frac{\mu^2}{M^2}\right).
\end{align}
The diagram $(b)$ in Fig.~\ref{fig:HHHH} gives
\begin{equation}
 i\mathcal{M}_{UV}^b = 3 \lambda_{\varphi H}^2 \mu^{4-d} 
 \int \frac{\mathrm{d}^{d}k}{(2\pi)^{d}} \frac{1}{D^2} 
 = 3 \lambda_{\varphi H}^2 \mu^{2\epsilon} A_2 
 = \frac{3 i \lambda_{\varphi H}^2}{16\pi^2} \log\frac{\mu^2}{M^2}\,.
\end{equation}
Finally,
\begin{equation}
 i\mathcal{M}_{UV} = i\mathcal{M}_{UV}^a + i\mathcal{M}_{UV}^b 
 =  \frac{2 i g_X^4}{\pi^2} + \frac{i}{16\pi^2} 
 \left(3 \lambda_{\varphi H}^2 - 12 g_X^4 \right) \log\frac{\mu^2}{M^2}\,.
 \label{eq:HHHH}
\end{equation}

\subsection*{Final remarks}
In light of the previous computations, we see that at the matching scale 
$\mu=M$, most of the one-loop corrections to renormalisable IR parameters 
vanish. The only exceptions are the Higgs mass parameter, kinetic term and 
quartic coupling. The relevant part of the IR Lagrangian reads
\begin{equation}
 \mathcal{L}_{SM+N}^{IR} \supset \alpha_H^{IR} \left(D_\mu H\right)^\dagger \left(D^\mu H\right) + 
 \left(\mu_H^{IR}\right)^2 H^\dagger H - \frac{1}{2}\lambda_H^{IR} \left(H^\dagger H\right)^2.
\end{equation}
(Note that in the UV the Higgs field is canonically normalised 
and therefore $\alpha_H^{UV}=1$.) Thus, the matching conditions read: 
\begin{equation}
i \alpha_H^{IR} p_h^2 + i \left(\mu_H^{IR}\right)^2 
= i p_h^2 + i \left(\mu_H^{UV}\right)^2 - i M^2_{UV}\,,
\end{equation}
where $-i M^2_{UV}$ is given in Eq.~\eqref{eq:MsqUV} and 
\begin{equation}
-3i\lambda_H^{IR} = -3i\lambda_H^{UV} + i \mathcal{M}_{UV}\,,
\end{equation}
with $i\mathcal{M}_{UV}$ from Eq.~\eqref{eq:HHHH}.
At the matching scale $\mu = M$ they lead to
\begin{equation}
 \alpha_H^{IR} = 1-\frac{g_X^2}{12\pi^2}\,, \quad
 \left(\mu_H^{IR}\right)^2 = \left(\mu_H^{UV}\right)^2
 +\frac{\lambda_{\varphi H}-4g_X^2}{16\pi^2}M^2\,, \quad
 \lambda_H^{IR} = \lambda_H^{UV} - \frac{2g_X^4}{3\pi^2}\,.
\end{equation}
%

\section{Matching the $\nu$SMEFT onto the $\nu$LEFT and anomalous dimensions}
\label{app:left}
%
The full list of lepton number conserving dimension-six operators in the 
$\nu$LEFT involving $N$ 
is shown in Tab.~\ref{tab:fermionic2}. 
(Those not involving $N$ can be found in Ref.~\cite{Jenkins:2017jig}.) 
\begin{table}[t]
\renewcommand{\arraystretch}{1.5}
\centering
\begin{tabular}{|c c c|}
\hline
Dipole&\multicolumn{2}{c|}{$\mathcal{O}_{N\gamma} = \overline{\nu_L}\sigma^{\mu\nu} N A_{\mu\nu}$}\\
\hline
\multirow{3}{*}{\vtext{RRRR}} & 
\multicolumn{2}{c|}{$\mathcal{O}_{NN}^{V,RR}=(\overline{N}\gamma_\mu N)(\overline{N}\gamma^\mu N)$} \\
&
 ${\cal O}_{eN}^{V,RR}=(\overline{e_R}\gamma_\mu e_R)(\overline{N}\gamma^\mu N)$ &
 ${\cal O}_{uN}^{V,RR}=(\overline{u_R}\gamma_\mu u_R)(\overline{N}\gamma^\mu N)$ \\
&
 ${\cal O}_{dN}^{V,RR}=(\overline{d_R}\gamma_\mu d_R)(\overline{N}\gamma^\mu N)$ &
${\cal O}_{udeN}^{V,RR}=(\overline{u_R}\gamma_\mu d_R)(\overline{e_R}\gamma^\mu N)$ \\
\hline
\multirow{3}{*}{\vtext{LLRR}} & 
 ${\cal O}_{\nu N}^{V,LR}=(\overline{\nu_L}\gamma_\mu \nu_L)(\overline{N}\gamma^\mu N)$ &
 ${\cal O}_{eN}^{V,LR}=(\overline{e_L}\gamma_\mu e_L)(\overline{N}\gamma^\mu N)$ \\
&
 ${\cal O}_{uN}^{V,LR}=(\overline{u_L}\gamma_\mu u_L)(\overline{N}\gamma^\mu N)$ &
 ${\cal O}_{dN}^{V,LR}=(\overline{d_L}\gamma_\mu d_L)(\overline{N}\gamma^\mu N)$ \\
&\multicolumn{2}{c|}{${\cal O}_{udeN}^{V,LR}=(\overline{u_L}\gamma_\mu d_L)(\overline{e_R}\gamma^\mu N)$} \\
\hline
\multirow{5}{*}{\vtext{LRLR}} & 
\multicolumn{2}{c|}{${\cal O}_{NN}^{S, RR}=(\overline{\nu_L} N) (\overline{\nu_L}N)$}\\
&
 ${\cal O}_{eN}^{S,RR}=(\overline{e_L} e_R)(\overline{\nu_L} N)$ &
 ${\cal O}_{eN}^{T,RR}=(\overline{e_L}\sigma_{\mu\nu}e_R) (\overline{\nu_L}\sigma^{\mu\nu} N)$ \\
&
 ${\cal O}_{uN}^{S,RR}=(\overline{u_L}u_R)(\overline{\nu_L}N)$ &
 ${\cal O}_{uN}^{T,RR}=(\overline{u_L}\sigma_{\mu\nu}u_R)(\overline{\nu_L}\sigma^{\mu\nu}N)$ \\
&
 ${\cal O}_{dN}^{S,RR}=(\overline{d_L}d_R)(\overline{\nu_L}N)$ &
 ${\cal O}_{dN}^{T,RR}=(\overline{d_L}\sigma_{\mu\nu}d_R)(\overline{\nu_L}\sigma^{\mu\nu}N)$ \\
&
${\cal O}_{udeN}^{S,RR}=(\overline{u_L}d_R)(\overline{e_L}N)$ &
${\cal O}_{udeN}^{T,RR}=(\overline{u_L}\sigma_{\mu\nu}d_R)(\overline{e_L}\sigma^{\mu\nu}N)$ \\
\hline
\multirow{2}{*}{\vtext{RLLR}} & 
${\cal O}_{eN}^{S,LR}=(\overline{e_R}e_L)(\overline{\nu_L}N)$ &
${\cal O}_{uN}^{S,LR}=(\overline{u_R}u_L)(\overline{\nu_L}N)$ \\
&
${\cal O}_{dN}^{S,LR}=(\overline{d_R}d_L)(\overline{\nu_L}N)$ &
${\cal O}_{udeN}^{S,LR}=(\overline{u_R}d_L)(\overline{e_L}N)$ \\
\hline
\end{tabular}
\caption{\it List of $\nu$LEFT lepton number conserving operators involving $N$. The addition of $h.c.$ is implied when needed. Note that the operators obtained from the non-Hermitian operators in the table with multiplication by the imaginary unit are also present. The notation follows that of Ref.~\cite{Jenkins:2017jig}, although we have not tried to minimise the number of operators involving $\sigma_{\mu\nu}$.}
\label{tab:fermionic2}
\end{table}
The following relations hold at the EW matching scale (note that we are ignoring family indices):
\begin{align}
 \frac{\alpha_{N\gamma}}{v} &= \frac{v}{\sqrt{2}\Lambda^2}\left(\alpha_{NB}c_W + \alpha_{NW} s_W\right), \Label{eq:match1} &
  \frac{\alpha_{NN}^{V, RR}}{v^2} &= \frac{\alpha_{NN}}{\Lambda^2}\,, \\
 \frac{\alpha_{eN}^{V, RR}}{v^2} &= \frac{\alpha_{eN}}{\Lambda^2} -\frac{g_Z^2 Z_{e_R} Z_N}{m_Z^2} \,, \Label{} &
 \frac{\alpha_{uN}^{V, RR}}{v^2} &= \frac{\alpha_{uN}}{\Lambda^2} -\frac{g_Z^2 Z_{u_R} Z_N}{m_Z^2}\,, \\
 \frac{\alpha_{dN}^{V, RR}}{v^2} &= \frac{\alpha_{dN}}{\Lambda^2} -\frac{g_Z^2 Z_{d_R} Z_N}{m_Z^2}\,, \Label{} &
 \frac{\alpha_{udeN}^{V,RR}}{v^2} &= \frac{\alpha_{duNe}}{\Lambda^2}\,, \\
 \frac{\alpha_{\nu N}^{V, LR}}{v^2} &= \frac{\alpha_{LN}}{\Lambda^2} -\frac{g_Z^2 Z_{\nu_L}Z_N}{m_Z^2}\,, \Label{} &
 \frac{\alpha_{eN}^{V, LR}}{v^2} &= \frac{\alpha_{LN}}{\Lambda^2} -\frac{g_Z^2 Z_{e_L}Z_N}{m_Z^2}\,, \\
 \frac{\alpha_{uN}^{V, LR}}{v^2} &= \frac{\alpha_{QN}}{\Lambda^2} -\frac{g_Z^2 Z_{u_L}Z_N}{m_Z^2}\,, \Label{} &
 \frac{\alpha_{dN}^{V, LR}}{v^2} &= \frac{\alpha_{QN}}{\Lambda^2} -\frac{g_Z^2 Z_{d_L}Z_N}{m_Z^2}\,, \\
 \frac{\alpha_{udeN}^{V, LR}}{v^2} &= -\frac{g^{2} W_N}{2m_W^2}\,, \Label{} &
  \alpha_{NN}^{S, RR} &= 0\,,\\
 \frac{\alpha_{eN}^{S, RR}}{v^2} &= \frac{3\alpha_{LNLe}}{2\Lambda^2}\,, \Label{} &
 \frac{\alpha_{eN}^{T, RR}}{v^2} &= \frac{\alpha_{LNLe}}{8\Lambda^2}\,, \\
 \alpha_{uN}^{S, RR} &= 0\,, \Label{} &
 \alpha_{uN}^{T, RR} &= 0\,, \\
 \frac{\alpha_{dN}^{S, RR}}{v^2} &= \frac{\alpha_{LNQd}}{\Lambda^2} - 
 \frac{\alpha_{LdQN}}{2\Lambda^2}\,, \Label{} &
 \frac{\alpha_{dN}^{T, RR}}{v^2} &= -\frac{\alpha_{LdQN}}{8\Lambda^2}\,, \\
 \frac{\alpha_{udeN}^{S, RR}}{v^2} &= \frac{\alpha_{LdQN}}{2\Lambda^2} - \frac{\alpha_{LNQd}}{\Lambda^2}\,, \Label{} &
 \frac{\alpha_{udeN}^{T, RR}}{v^2} &= \frac{\alpha_{LdQN}}{8\Lambda^2}\,, \\
 \frac{\alpha_{eN}^{S, LR}}{v^2} &= \frac{g^2 W_N}{m_W^2}\,, \Label{} &
 \frac{\alpha_{uN}^{S, LR}}{v^2} &= \frac{\alpha_{QuNL}}{\Lambda^2}\,, \\
 \alpha_{dN}^{S, LR} &=0\,, \Label{} &
 \frac{\alpha_{udeN}^{S, LR}}{v^2} &= \frac{\alpha_{QuNL}}{\Lambda^2}\,. \label{eq:match2}
\end{align}
The coupling $g_Z$ is defined as $g_Z = e/(s_W c_W)$. We have also defined $Z_{\psi_{SM}} = T_3-Q s_W^2$ and $Z_N = -\alpha_{HN} v^2/(2\Lambda^2)$ as well as $W_N = \alpha_{HNe} v^2/(2\Lambda^2)$. Note that we can neglect EFT effects in the non $N$ fermion couplings to the $Z$ and $W$ because they would lead to dimension-eight contributions. 

\begin{figure}[t]
 \centering
  \includegraphics[width=0.5\textwidth]{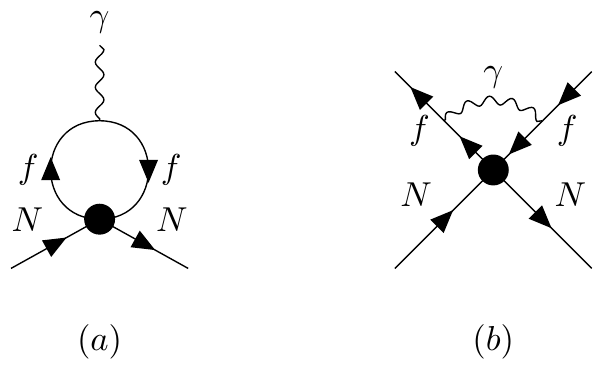}
 \caption{\it $(a)$ Renormalisation of the operator $\partial^\nu A_{\mu\nu}\overline{N}\gamma^\mu N$ by four-fermions. It generates other four-fermions upon using the equations of motion. $(b)$ Self-renormalisation of four-fermions.}
 \label{fig:running}
\end{figure}

In our case, the only operators that are generated are the dipole as well as 
vector type RR and LR four-fermions with two $N$s. They renormalise due to quantum 
corrections depicted by the diagrams in Fig.~\ref{fig:running}. 
Using the notation 
\begin{equation}
 \dot{\alpha} \equiv 16\pi^2\mu \frac{\mathrm{d}\alpha}{\mathrm{d}\mu}\,,
\end{equation}
we obtain: 
\begin{equation}
 \dot{\alpha}_{N\gamma} = \frac{4}{3}\left(3q_e^2 + 3N_c q_d^2 + 2N_c q_u^2\right) e^2\alpha_{N\gamma}\,,
\end{equation}
\begin{align}
 \dot{\alpha}_{\psi N}^{V,RR} = \frac{4}{3}e^2 q_\psi\bigg[N_c q_u \left(\alpha_{uN}^{V,RR} + \alpha_{uN}^{V,LR}\right) &+ N_c q_d \left(\alpha_{dN}^{V,RR}+\alpha_{dN}^{V,LR}\right) \nonumber\\
 &+ q_e\left(\alpha_{eN}^{V,RR} + \alpha_{eN}^{V,LR}\right)\bigg]\,,
\end{align}
\begin{align}
 \dot{\alpha}_{\psi N}^{V,LR} = \frac{4}{3}e^2 q_\psi\bigg[N_c q_u \left(\alpha_{uN}^{V,RR} + \alpha_{uN}^{V,LR}\right) &+ N_c q_d \left(\alpha_{dN}^{V,RR}+\alpha_{dN}^{V,LR}\right) \nonumber\\
 &+ q_e\left(\alpha_{eN}^{V,RR} + \alpha_{eN}^{V,LR}\right)\bigg]\,,
\end{align}
for $\psi = \nu, N, e, u, d$. The non-vanishing electric charges are $q_e = -1$, $q_u = 2/3$ and $q_d=-1/3$. This automatically implies that $\dot{\alpha}_{NN}^{V,RR} = 0$ and $\dot{\alpha}_{\nu N}^{V,LR} = 0$; \textit{i.e.} these operators do not renormalise.

Finally, lepton number violating operators are also induced within 
our framework when $m_N\neq 0$. 
There are 19 operators violating lepton number by two units 
and one violating it by four units.
We list them in Tab.~\ref{tab:fermionic3}.
\begin{table}[t]
\renewcommand{\arraystretch}{1.5}
\centering
\begin{tabular}{|c c c|}
\hline
\multirow{3}{*}{\vtext{LLLL}} & 
 ${\cal O}_{\nu N^c}^{V,LL}=(\overline{\nu_L}\gamma_\mu \nu_L)(\overline{\nu_L}\gamma^\mu N^c)$ & 
 ${\cal O}_{eN^c}^{V,LL}=(\overline{e_L}\gamma_\mu e_L)(\overline{\nu_L}\gamma^\mu N^c)$ \\
& 
 ${\cal O}_{uN^c}^{V,LL}=(\overline{u_L}\gamma_\mu u_L)(\overline{\nu_L}\gamma^\mu N^c)$ &
 ${\cal O}_{dN^c}^{V,LL}=(\overline{d_L}\gamma_\mu d_L)(\overline{\nu_L}\gamma^\mu N^c)$ \\
 & \multicolumn{2}{c|}{${\cal O}_{udeN^c}^{V,LL}=(\overline{u_L}\gamma_\mu d_L)(\overline{e_L}\gamma^\mu N^c)$} \\
\hline
\multirow{2}{*}{\vtext{RRLL}} & 
 ${\cal O}_{eN^c}^{V,RL}=(\overline{e_R}\gamma_\mu e_R)(\overline{\nu_L}\gamma^\mu N^c)$ &
 ${\cal O}_{uN^c}^{V,RL}=(\overline{u_R}\gamma_\mu u_R)(\overline{\nu_L}\gamma^\mu N^c)$ \\
&
 ${\cal O}_{dN^c}^{V,RL}=(\overline{d_R}\gamma_\mu d_R)(\overline{\nu_L}\gamma^\mu N^c)$ & 
 ${\cal O}_{udeN^c}^{V,RL}=(\overline{u_R}\gamma_\mu d_R)(\overline{e_L}\gamma^\mu N^c)$ \\
\hline
\multirow{3}{*}{\vtext{RLRL}} 
&
 ${\cal O}_{eN^c}^{S,LL}=(\overline{e_R} e_L)(\overline{N} N^c)$ & 
 ${\cal O}_{uN^c}^{S,LL}=(\overline{u_R}u_L)(\overline{N}N^c)$ \\
&
 ${\cal O}_{dN^c}^{S,LL}=(\overline{d_R}d_L)(\overline{N}N^c)$ & 
 ${\cal O}_{udeN^c}^{S,LL}=(\overline{u_R}d_L)(\overline{e_R}N^c)$ \\
& 
\multicolumn{2}{c|}{${\cal O}_{udeN^c}^{T,LL}=(\overline{u_R}\sigma_{\mu\nu}d_L)(\overline{e_R}\sigma^{\mu\nu}N^c)$} \\
\hline
\multirow{3}{*}{\vtext{LRRL}} 
&
 ${\cal O}_{\nu^cN^c}^{S,RL}=(\overline{\nu_L}\nu_L^c)(\overline{N}N^c)$ & 
 ${\cal O}_{NN^c}^{S,RL}=(\overline{\nu_L}N)(\overline{N}N^c)$ \\
&
 ${\cal O}_{eN^c}^{S,RL}=(\overline{e_L}e_R)(\overline{N}N^c)$ &
 ${\cal O}_{uN^c}^{S,RL}=(\overline{u_L}u_R)(\overline{N}N^c)$ \\
&
 ${\cal O}_{dN^c}^{S,RL}=(\overline{d_L}d_R)(\overline{N}N^c)$ &
 ${\cal O}_{udeN^c}^{S,RL}=(\overline{u_L}d_R)(\overline{e_R}N^c)$ \\
\hline
\end{tabular}
\caption{\it List of $\nu$LEFT lepton number violating 
operators involving $N^c$. The addition of $h.c.$ is implied when needed. 
Note that the operators obtained from the non-Hermitian operators  
in the table with multiplication by the imaginary unit are also present.}
\label{tab:fermionic3}
\end{table}
(The subset of these operators relevant for neutrinoless double beta decay has 
been recently provided in Ref.~\cite{Dekens:2020ttz}, which also considers 
higher dimensional operators, the running down to the QCD scale as well as the 
matching onto the chiral perturbation theory.)

If $m_N$ is not vanishing, the $\nu$SMEFT operators $\mathcal{O}_{NW}$ 
and $\mathcal{O}_{NB}$ induce, after EWSB, lepton number violating operators proportional to $m_N 
v/\Lambda^2$ upon using the equation of motion
$i\slashed{\partial}N\sim m_N N^c$. These are: $\mathcal{O}_{\nu N^c}^{V,LL}$, 
$\mathcal{O}_{e N^c}^{V,LL}$, $\mathcal{O}_{e N^c}^{V, RL}$, 
$\mathcal{O}_{uN^c}^{V,LL}$, $\mathcal{O}_{uN^c}^{V,RL}$, 
$\mathcal{O}_{dN^c}^{V,LL}$ and $\mathcal{O}_{dN^c}^{V,RL}$ ($Z$ mediated) as 
well as $\mathcal{O}_{udeN^c}^{V,LL}$ and $\mathcal{O}_{eN^c}^{V,LL}$ 
($W$ mediated).

\bibliography{vSMEFT_matching_v2}

\end{document}